\documentclass[11pt]{article}
\pdfoutput=1
\usepackage{draft}
\usepackage{comment}
\usepackage[normalem]{ulem}
\usepackage{hyperref}
\hypersetup{
 colorlinks=true,
 urlcolor=blue,
 anchorcolor=blue,
 citecolor=blue,
 filecolor=blue,
 linkcolor=blue,
 menucolor=blue,
 pagecolor=blue,
 linktocpage=true,
}
\usepackage{graphicx,color,subfig}
\usepackage{cite}
\usepackage{empheq}
\usepackage[font=small,labelfont=bf]{caption} 

\usepackage{bbold}
\usepackage[usenames,dvipsnames,table]{xcolor}
\graphicspath{{./figures/}}
\usepackage{tikz}

\newcommand{\ZZ}{\mathbb{Z}}
\newcommand{\RR}{\mathbb{R}}

\newcommand{\id}{\mathbb{1}}

\newcommand{\half}{{1\over 2}}

\DeclareMathOperator{\im}{Im}
\DeclareMathOperator{\re}{Re}
\DeclareMathOperator{\rank}{rank}
\DeclareMathOperator{\diag}{diag}
\DeclareMathOperator{\Vol}{Vol}
\newcommand{\mymod}{\,{\rm mod}\,}

\DeclarePairedDelimiter\floor{\lfloor}{\rfloor}
\newcommand\ofone{{}_0F_1}

\newcommand\Z{{\mathbb Z}}
\newcommand\R{{\mathbb R}}
\newcommand\mat[4]{\begin{pmatrix}#1 & #2 \\ #3 & #4 \end{pmatrix}}
\def\be{\begin{equation}}
\def\ee{\end{equation}}

\usepackage{scalerel}
\usepackage{stackengine,wasysym}

\newcommand\reallywidetilde[1]{\ThisStyle{%
  \setbox0=\hbox{$\SavedStyle#1$}%
  \stackengine{-.1\LMpt}{$\SavedStyle#1$}{%
    \stretchto{\scaleto{\SavedStyle\mkern.2mu\AC}{.5150\wd0}}{.6\ht0}%
  }{O}{c}{F}{T}{S}%
}}

\DeclareFontFamily{OT1}{pzc}{}
\DeclareFontShape{OT1}{pzc}{m}{it}{<-> s * [1.10] pzcmi7t}{}
\DeclareMathAlphabet{\mathpzc}{OT1}{pzc}{m}{it}

\definecolor{vert}{rgb}{0.1367 0.543 0.1367}

\def\({\left(}
\def\){\right)}

\begin{document}

\unitlength = .8mm

\begin{titlepage}

	%  \begin{flushright}
	% \hfill{\tt PUPT-\sac{}}
	% \end{flushright} 

\begin{center}

 \hfill \\
 \hfill \\

\title{Wormholes and Spectral Statistics in the Narain Ensemble}

\author{Scott Collier$^{a}$ and Alexander Maloney$^{b}$}

 \address{
$^{a}$ Princeton Center for Theoretical Science, Princeton University, 
Princeton, NJ 08544, USA
\\
 $^{b}$ Department of Physics, McGill University,
 Montreal, QC H3A 2T8, Canada
 }

\email{scott.collier@princeton.edu, alex.maloney@mcgill.ca}

\end{center}

\abstract{
We study the spectral statistics of primary operators in the recently formulated ensemble average of Narain's family of free boson conformal field theories, which provides an explicit (though exotic) example of an averaged holographic duality. In particular we study moments of the partition function by explicit computation of higher-degree Eisenstein series.  This describes the analog of wormhole contributions coming from a sum of over geometries in the dual theory of ``U(1) gravity" in AdS$_3$. We give an exact formula for the two-point correlation function of the density of primary states.  We compute the spectral form factor and show that the wormhole sum reproduces precisely the late time plateau behaviour related to the discreteness of the spectrum.  The spectral form factor does not exhibit a linear ramp.
}

\vfill

\end{titlepage}

\eject

\begingroup

\tableofcontents

\endgroup

\section{Introduction}\label{sec:intro}

In the study of the quantum mechanics of black holes, an important challenge is to find models that are simple enough that they can be studied in an exact manner, but sophisticated enough to capture the rich dynamics of realistic black holes. For example, supersymmetric black holes in string theory are, while far from generic, important examples whose states can be precisely enumerated and matched with the semiclassical Bekenstein-Hawking entropy \cite{Strominger:1996sh}. But black holes are dynamical objects which are characterized by more than just an entropy: they are conjectured to be the fasted scramblers of information in nature \cite{Sekino:2008he}, and are expected to exhibit rich chaotic dynamics \cite{Shenker:2013pqa,Maldacena:2015waa,Cotler:2016fpe}.  To study the chaotic dynamics of black holes, an important class of examples are provided by the AdS/CFT correspondence, where the high energy states of a (typically strongly coupled) CFT are dual to the quantum states of a black hole.

In seeking solvable examples it is natural to focus on AdS$_3$/CFT$_2$ where the boundary theory has an infinite number of symmetries that aid in solvability.  Indeed, in this case Cardy's formula \cite{Cardy:1986ie} for the high energy density of states exactly matches the semiclassical entropy of the corresponding black holes in AdS$_3$ \cite{Strominger:1997eq}. A CFT$_2$ possesses an infinite number of symmetries generated by some number $c_{\rm currents}$ of chiral operators.  In a generic CFT the chiral algebra is generated by the stress tensor so $c_{\rm currents}=1$, but many special CFTs have $c_{\rm currents}>1$.\footnote{We will focus in this paper on parity-invariant CFTs, so that the theory will have both $c_{\rm currents}$ chiral and $c_{\rm currents}$ anti-chiral currents.}  As $c_{\rm currents}$ increases the theory becomes, in a sense, more solvable.  So it is natural to ask: how large can one make $c_{\rm currents}$ and still have theory which, in an appropriate sense, describes black hole dynamics in the bulk?
A hint is provided by Cardy's formula
\be
\label{cardy}
\log \rho^p(\Delta) \sim 4\pi \sqrt{\frac{c-c_{\rm currents}}{24} \Delta}+ n \log \Delta + \dots 
\ee
for the density of primary states with large scaling dimension $\Delta$, where $n$ is a non-universal constant.\footnote{For brevity we write here only the formula for spinless states. States with spin will be considered in great detail later.}  Here, by primary states we are referring to states that are primary with respect to the full chiral algebra: they are annihilated by all of the lowering operators constructed out of the chiral algebra.  We focus on primary states because chiral operators are dual to gauge fields in the bulk; the action of a raising operator simply dresses a bulk state with the edge modes of a bulk gauge field. For example, applying a Virasoro raising operator to a state will dress it with a so-called ``boundary graviton" associated to large diffeomorphisms that act on the boundary of AdS$_3$. In this sense the chiral algebra describes the dynamics of edge modes that live at the boundary. To study the dynamics of black holes in the bulk we must therefore focus on primary states.

At fixed $c_{\rm currents}$, the semiclassical holographic dictionary relates the central charge of the asymptotic symmetry algebra to the AdS$_3$ radius via \cite{Brown:1986nw}
\begin{equation}
	c = {3\ell\over 2G_N} + \mathcal{O}(1)~.
\end{equation} 
The leading term in the Cardy formula (\ref{cardy}) is proportional to the area of the black hole, so can be interpreted as the usual Bekenstein-Hawking entropy. Something clearly goes wrong if $c<c_{\rm currents}$.  In this case the theory is expected to have only a finite number of operators that are primary with respect to the full chiral algebra, and is typically a rational theory whose correlation functions are completely determined by Ward identities of the chiral algebra. Thus when $c<c_{\rm currents}$ there seems to be no sense in which the high energy dynamics of the theory is dominated by states that resemble semi-classical black holes.\footnote{Though there may nevertheless be a sense in which it is possible to interpret such theories as exotic theories of gravity; see \cite{Castro:2011zq,Jian:2019ubz} for examples with Virasoro symmetry and $c<1$ where the partition function can be expressed as a sum over modular images which one may attempt to interpret as a sum over geometries, and \cite{Gaberdiel:2010pz} for examples with ${\cal W}_N$ symmetry and $c<N-1$.} When $c>c_{\rm currents}$, on the other hand, the theory is expected to be irrational and to exhibit chaotic dynamics much like CFTs in higher dimensions. Although some universal properties of the spectrum of local operators and their dynamics in such theories are known \cite{Hellerman:2009bu,Friedan:2013cba,Collier:2016cls,Afkhami-Jeddi:2019zci,Hartman:2019pcd,Kusuki:2018wpa,Collier:2018exn,Collier:2019weq}, they seem to be nearly as difficult to study as their higher dimensional cousins. 

This leaves open the question of theories with $c=c_{\rm currents}$. In such theories the number of primary states typically grows polynomially with dimension rather than exponentially, with a power that depends on the theory that we have called $n$ in equation (\ref{cardy}).  Indeed, the Bekenstein-Hawking formula typically contains a correction of the form $S = {A \over 4G_N} + n \log A + \dots$.  So one might hope that in such theories the primary states could be identified with black holes whose leading semi-classical entropy vanishes for some reason, but whose entropy is nevertheless large due to the subleading term.\footnote{There is an interesting historical precedent for this.  Prior to Strominger and Vafa's computation of black hole entropy \cite{Strominger:1996sh}, Sen \cite{Sen:1995in} investigated a simpler class of D-brane bound states (with fewer non-zero charges) which describe black holes with vanishing horizon area but non-zero entropy. Such configurations can be viewed as black holes with string-scale stretched horizons, which have non-zero area only once certain quantum effects are included (as discussed in e.g. \cite{Dabholkar:2004yr,Dabholkar:2004dq,Sen:2004dp,Hubeny:2004ji,Dabholkar:2005dt,Benjamin:2016aww}).} One might hope, then, that these are the Goldilocks CFTs: simple enough to solve, but complex enough to provide a useful model for black hole dynamics.

One is thus led to ask to what extent the primary states of CFTs with $c=c_{\rm currents}$ resemble black holes, even absent a semi-classical Bekenstein-Hawking entropy proportional to area. Our focus will be on the statistical properties of the spectrum of such states, and on whether they exhibit the spectral statistics characteristic of a chaotic system. We will consider the simplest possible set of such theories, those with $U(1)^D \times U(1)^D$ current algebra and $c=D$. It is believed\footnote{But not, to the best of our knowledge, proven.} that the most general compact, unitary, and modular invariant theory with this current algebra and central charge is a theory of $D$ free bosons, a sigma model with a $D$-dimensional torus as target space.  This can be viewed as the worldsheet theory of a toroidal compactification of string theory.  Such CFTs are not unique, and are parameterized by a moduli space first described by Narain \cite{Narain:1985jj,Narain:1986am}.  The data required to specify such a CFT is an even self-dual lattice of signature $(D,D)$, which is parameterized by a $D^2$ dimensional moduli space 
\be
{\cal M}_D = O(D,D;\Z)\backslash O(D,D;\R)/O(D)\times O(D)
\ee 
In the language of the sigma model with toroidal target space, the moduli space ${\cal M}_D$ is parameterized by the space of metrics and $B$ field fluxes on the $D-$dimensional torus.

In this paper, we are interested in exploring the properties of primary states of these theories. The spectrum of a conformal field theory is captured by its torus partition function, which is a thermal trace over the Hilbert space of states of the CFT on the circle\footnote{For the sake of brevity we will write partition functions in this paper as $Z(\tau)$ rather than $Z(\tau,\bar\tau)$.  But we emphasize that these are not holomorphic quantities, and we are not studying chiral CFTs.}
\begin{equation}
	Z(\tau)= \Tr_{\mathcal{H}_{S^1}}\left(q^{L_0-c/24} {\bar q}^{{\widetilde L}_0-c/24}\right),\quad q = e^{2\pi i \tau}
\end{equation}
where $\tau$ is the modulus of the torus. The spectrum of the Narain theories is organized into representations of the $U(1)^D\times U(1)^D$ chiral algebra. We are particularly interested in the partition function that counts only states that are primary with respect to the full chiral algebra, which we can write as\footnote{Note that the partition function $Z^p(\tau)$ defined this way is not modular invariant but rather transforms with weight $({D\over 2},{D\over 2})$ under $SL(2,\mathbb{Z})$.}
\begin{equation}
	Z^p(\tau) = |\eta(\tau)|^{2D}Z(\tau).
\end{equation}
The Cardy formula for the asymptotic density of (scalar) primary states is given by
\begin{equation}\label{cardyy}
	\log\rho^p(\Delta) \sim (D-2) \log \Delta + \ldots
\end{equation}
Although numerically large, this is much smaller than the density of descendant states at high energy, which exhibits the usual Cardy growth that is exponential in the square-root of $\Delta$ when $\Delta$ is much larger than the central charge. In the gravitational language, this reflects the fact that most of the energy of a typical heavy state would come from a gas of ``boundary photons" associated with the gauge fields dual to the $U(1)$ currents, with only a relatively small amount coming from the primary state; in this sense the Narain theories are quite different from the duals of semiclassical gravity in AdS$_3$, where the asymptotic densities of both primary and descendant states grow exponentially (with a larger exponent for the primary states). Nevertheless, aside from the fact that the entropy scales like $\log \Delta$ rather than $\sqrt{\Delta}$, equation (\ref{cardyy}) has many features which resemble a traditional Bekenstein-Hawking formula. For example, the entropy grows linearly with the central charge $D$, which still plays the role of the coupling constant in the bulk theory.  The $\dots$ are all subleading in the large $D$ limit, so one might expect (as will indeed be the case) that in this limit the primary-counting partition function exhibits a Hawking-Page phase transition much like any other theory of semi-classical gravity in AdS$_3$.  Indeed, as we will review below, in the gravitational interpretation of the Narain ensemble the leading asymptotic growth (\ref{cardy}) comes from a saddle point which is precisely the Euclidean BTZ black hole.

To probe the spectral statistics of primary operators in the Narain family of theories, we would like to go beyond the density of states (\ref{cardyy}). One might expect that the simplest way to proceed would be to pick some particular theories from the Narain family and study their spectral statistics. However, it turns out to be both more interesting, and significantly simpler, to pursue a different strategy. We first note that, as ${\cal M}_D$ is a group manifold it has a unique invariant measure $d\mu$ --- the Haar measure --- that coincides with Zamolodchikov's measure on the moduli space of CFTs.  Thus we can average over this moduli space, and hence investigate notions of what it means to be a ``typical" CFT in this moduli space with respect to this measure.  In particular, for any function on moduli space we define the average
\be
	\langle \cdot \rangle = {1\over {\Vol({\cal M}_D)}} \int_{{\cal M}_D} d\mu \left(\cdot\right). 
\ee
To study the spectral statistics of Narain CFTs we will compute correlation functions of the form
\be
	\langle Z^p(\tau_1)\dots Z^p(\tau_n)\rangle
\ee
We note that, even though each instance of the Narain ensemble is integrable (at least in the sense that $c=c_{\rm currents}$) there is some precedent for the idea that an ensemble of integrable quantum mechanical systems may nevertheless exhibit interesting properties: the quadratic SYK model \cite{Winer:2020mdc}, which develops an exponential ramp in the spectral form factor, is a recently studied example.\footnote{We are grateful to Douglas Stanford for bringing this to our attention.} That it is easier to study these averages than to study a particular typical instance of the Narain ensemble mirrors a similar phenomenon that occurs in the sphere packing problem.\footnote{The analogy between two dimensional CFTs and sphere packing was made quite sharp in \cite{Hartman:2019pcd}.}

Another motivation for the study of an ensemble of CFTs is due to remarkable recent developments which suggest that some quantum gravitational systems are dual to an ensemble average of many quantum theories rather than to a particular quantum mechanical system. An early hint of this came from the Sachdev-Ye-Kitaev model \cite{Sachdev:1992fk,Kitaev:2015aa,Maldacena:2016hyu}, a one-dimensional model of Majorana fermions with a random interaction that in the infrared develops an approximate conformal symmetry and saturates the chaos bound \cite{Maldacena:2015waa}, and thus provides a useful model for nearly-AdS$_2$ holography. Later, it was shown that its low-energy limit, Jackiw-Teitelboim gravity \cite{Jackiw:1984je,Teitelboim:1983ux,Almheiri:2014cka,Maldacena:2016upp}, admits a non-perturbative completion in terms of a double-scaled random matrix integral \cite{Saad:2019lba,Stanford:2019vob,Kapec:2019ecr}. More recently, there have been hints that the gravitational path integral of pure three-dimensional quantum gravity \cite{Maloney:2007ud,Keller:2014xba} may in some sense be dual to an ensemble of irrational conformal field theories (see e.g. \cite{Cotler:2020ugk,Cotler:2020hgz,Maxfield:2020ale}).\footnote{Although what exactly is meant by this has not yet been completely articulated, partly because the space of such CFTs is still poorly understood.} The consequences of this would be striking.  For example, it would mean that the typical large $c$ CFT has a lightest primary state with dimension $\Delta= {\cal O}(c)$. Nevertheless, such a duality would resolve two puzzles in the study of three dimensional gravity. The first is the existence of ``Euclidean wormhole" solutions of AdS$_3$ gravity which apparently lead to non-factorization of the partition function on disconnected surfaces \cite{Maldacena:2004rf}. Second, it would explain the continuous spectrum obtained from the explicit sum over gravitational instantons in Einstein gravity \cite{Maloney:2007ud,Keller:2014xba}.\footnote{This sum also suffers from an apparent breakdown of unitarity \cite{Maloney:2007ud,Keller:2014xba,Benjamin:2019stq}, which is particularly sharp in the near-extremal limit at large-spin \cite{Benjamin:2019stq}, although there are problems even at finite spin \cite{Alday:2019vdr}; proposed fixes include modifications of the gravitational path integral to include conical singularities \cite{Benjamin:2020mfz,Alday:2019vdr} (see also \cite{Alday:2020qkm} for a discussion of this in higher-spin AdS$_3$ gravity) or configurations that are not continuously connected to classical saddles \cite{Maxfield:2020ale}. The ensemble average of Narain theories does not suffer from this non-unitarity.}  It may, moreover, provide an interpretation for the universal formulas for asymptotic CFT data \cite{Kraus:2016nwo,Das:2017cnv,Brehm:2018ipf,Romero-Bermudez:2018dim,Hikida:2018khg,Collier:2018exn,Collier:2019weq} in terms of an ensemble average of CFTs rather than a microcanonical average over high-energy states in any particular CFT (see \cite{Belin:2020hea} for a related discussion).

This motivates the study of the ensemble average of Narain's family of free boson conformal field theories, with the hope that this will provide a concrete example of an ensemble of conformal field theories which is dual to a theory of three-dimensional gravity.  Of course, because these CFTs are all free we anticipate that the dual theory of gravity will, in an appropriate sense, be {\it simpler} than three dimensional Einstein gravity.\footnote{Indeed, the Narain ensemble is quite different from the ensemble of generic CFTs. Perhaps the most important difference is that Narain moduli space is a continuous conformal manifold connected by exactly marginal deformations.  An ensemble putatively dual to pure Einstein gravity will presumably include isolated points in addition, perhaps, to connected moduli spaces of CFTs.} Remarkably, averages over Narain moduli space can be computed using the Siegel-Weil formula \cite{Siegel_1951,Weil_1964,Weil_1965}.  The interpretation of these formulas in terms of a bulk theory of gravity was described in \cite{Maloney:2020nni,Afkhami-Jeddi:2020ezh}. It was noted in \cite{Maloney:2020nni} that for an arbitrary Riemann surface $\Sigma$ the result for the average can be interpreted as a sum over classical saddle points:
\be\label{Smiley}
\langle Z(\Sigma) \rangle = \sum_{g_0} e^{-D S^{(0)}[g_0] + S^{(1)}[g_0]}
\ee
When the surface $\Sigma$ is connected, the $g_0$ correspond to classical geometries (in particular handlebodies) with constant negative curvature that have the surface $\Sigma$ as their boundary.  The $S^{(0)}$ and $S^{(1)}$ are the classical and one-loop actions of these saddle points (where $D\sim \hbar^{-1}$ plays the role of the coupling constant) which can be computed explicitly in terms of a bulk Chern-Simons theory.  In a sense, equation (\ref{Smiley}) is evidence for a new holographic duality between our boundary ensemble and an exotic bulk theory of gravity, dubbed ``U(1) Gravity" \cite{Afkhami-Jeddi:2020ezh}, that includes $U(1)^{2D}$ Chern-Simons theory as its perturbative piece and a sum over handlebodies as its non-perturbative dynamics.\footnote{The gravitational description of the Narain ensemble was further studied in \cite{Perez:2020klz,Dymarsky:2020pzc,Datta:2021ftn}, and the discrete average of holomorphic free boson CFTs was discussed in \cite{Dymarsky:2020qom}. Aspects of ensemble averages over other classes of rational two-dimensional conformal field theories have also been discussed \cite{Meruliya:2021utr,Benjamin:2021wzr,Meruliya:2021lul,Ashwinkumar:2021kav,Dong:2021wot}.}

The primary states in the Narain ensemble should be understood as the analog of black hole microstates in this theory. To see this, one can imagine computing the average density of states by evaluating (\ref{Smiley}) when $\Sigma$ is a torus. The high energy behaviour of the density of states matches the Cardy formula (\ref{cardyy}), and comes from a saddle point which is precisely the Euclidean BTZ black hole.\footnote{In fact, the subleading behaviour in (\ref{cardyy}) -- and in particular the subleading term which scales like ${\cal O}(\Delta^0)$ -- requires the inclusion of the full ``$SL(2,{\Z})$'' family of Euclidean black holes.}   Our Narain theory therefore provides a solvable toy model were the properties of black hole microstates can be investigated explicitly. Our goal is to understand the statistical properties of these states.  

Our primary object of interest will be the correlation functions of the torus partition function, which --- according to the Siegel-Weil formula --- can be also written in the form:
\be\label{Karla}
\langle Z(\tau_1)\dots Z(\tau_n) \rangle = \sum_{g_0} e^{-D S^{(0)}[g_0] + S^{(1)}[g_0]}
\ee  
This sum will be our main object of interest in this paper.  It turns out to be what is known as a real analytic Eisenstein series of degree $n$, and its gravitational interpretation was initiated in \cite{Maloney:2020nni}. The $g_0$ which appear in this sum are the analogs of Euclidean wormholes, albeit for $U(1)$ gravity rather than for Einstein gravity. One important difference between this sum and equation (\ref{Smiley}) is that we can no longer interpret the $g_0$ as handlebodies with a constant negative curvature metric; indeed, they need not correspond to hyperbolic geometries at all.  Nevertheless, they are quite similar to the Euclidean wormholes which appear in three dimensional Einstein gravity.  For example, as described in \cite{Maloney:2020nni},  many of the terms in the sum can be associated to bulk topologies which connect $n$ boundary tori.  Indeed, we will see that it is precisely these contributions which dominate the sum in many cases.  For example, they lead to the famous ``plateau" in the spectral form factor.  Thus, although the analogy with Einstein gravity is perhaps not perfect, we will regard our Eisenstein series as a sum over Euclidean wormholes. 

We will begin by reviewing in section \ref{sec:holography} the Siegel-Weil formula and its holographic interpretation. In section \ref{sec:partitionFunction} we will review its simplest application: the computation of $\langle Z^p(\tau)\rangle$, the averaged torus partition function, which gives the averaged density of states $\langle \rho^p(\Delta)\rangle$.  This will provide us a useful warm-up to the central exercise of section \ref{sec:twoPoint}, the computation of the two-point function $\langle Z^p(\tau_1) Z^p(\tau_2)\rangle$.  We will discuss the interpretation of this two point function as coming from a sum over wormholes, and give an explicit expression for the density-density correlation function $\langle \rho^p(\Delta_1) \rho^p(\Delta_2)\rangle$. In section \ref{sec:higherPoint} we will discuss the computation of higher-point functions of the torus partition function.  In section \ref{sec:SFF} we will study the spectral form factor $\langle Z^p(\beta+iT) Z^p(\beta-iT)\rangle$, which provides a useful characterization of the spectral statistics of the Narain family of CFTs. We will see that our expression precisely reproduces the expected late-time ``plateau" in the spectral form factor. We will pay particular attention to the comparison with Einstein gravity, and end with a brief discussion. In appendix \ref{app:crossingKernels} we will provide an alternate derivation of the averaged one- and two-point functions of the density of states using modular crossing kernels, and in appendix \ref{app:eisensteinFourier} we provide technical (but important) details on the higher-degree Eisenstein series that are relevant for the computation of averaged higher-genus partition functions and of higher moments of the torus partition function.

\section{An ensemble of free conformal field theories and holographic duality}\label{sec:holography}
In this paper we will study the simplest class of two-dimensional conformal field theories with $c=c_{\rm currents}$, namely Narain's family of free CFTs. The spectrum of such theories is organized into representations of the $U(1)^D\times U(1)^D$ current algebra  with central charge $c=D$. The authors of \cite{Maloney:2020nni,Afkhami-Jeddi:2020ezh} initiated the study of the ensemble of Narain CFTs, in which observables are computed by averaging over the moduli space of such CFTs with a measure naturally determined by the Zamolodchikov metric. 

The data that determines a Narain CFT is an even self dual lattice $\Lambda$  in $\mathbb{R}^{D,D}$.  The partition function of the CFT on a Riemann surface can be written in terms of a nonholomorphic theta function associated with $\Lambda$. The theta function depends on the moduli of the Riemann surface through the period matrix $\Omega$, which is an element of the Siegel upper half-space $\mathcal{H}_g$ (appendix \ref{app:eisensteinFourier} includes a brief review of Siegel upper-half space). The theta function also depends on the data of the lattice, which we can parametrize through the metric $G_{ab}$ and $B$-field $B_{ab}$ of the toroidal target space.  We will collectively denote these target space moduli by $m$.  The theta function is
\begin{equation}
\begin{aligned}
  \Theta(\Omega,m) &= \sum_{n_a^{~i},w_b^{~j}\in\mathbb{Z}^{D\times g}}\exp\left[-{\pi\over \alpha'}Y_{ij}\left(G^{ab}v_a^{~i}v_b^{~j}+G_{cd}w^{ci}w^{dj}\right)+2\pi i X_{ij}\delta^a_c n_a^{~i}w^{cj}\right],
\end{aligned}
\end{equation}
where $X$ and $Y$ respectively label the real and imaginary parts of $\Omega$,
\begin{equation}
\begin{aligned}
	v_a^{~i} &= \alpha' n_a^{~i} + B_{ac}w^{ci}
\end{aligned}
\end{equation}
and $n$ and $w$ are $D\times g$ matrices with integer entries that label momentum and winding modes of the free boson. The combination $(\det Y)^{D\over 2}\Theta(\Omega,m)$ is invariant under the action of the modular group $Sp(2g,\mathbb{Z})$ (which we describe in appendix \ref{app:eisensteinFourier}) on the period matrix.

The partition function is equal to $\Theta(\Omega,m)$ times a prefactor factor which is independent of $m$.  
One can think of the theta function as describing the contribution of the $U(1)^D \times U(1)^D$ primary states to the partition function, and this prefactor as describing the contribution of descendant states.
This is easiest to see for the torus partition function, which is
\begin{equation}\label{eq:latticePartitionFunction}
	Z(\tau,m) = {\Theta(\tau,m)\over |\eta(\tau)|^{2D}},
\end{equation}
where $\tau$ is the modular parameter of the torus. The Dedekind eta functions in the denominator of (\ref{eq:latticePartitionFunction}) enumerate the descendants with respect to the $U(1)^D\times U(1)^D$ chiral algebra. The spectrum of local operators that are primary with respect to the full chiral algebra are determined by the momentum and winding numbers and moduli $m$, with the dimension and spin given respectively by
\begin{equation}
\begin{aligned}
	\Delta &= {1\over2\alpha'}\left(G^{ab}v_a v_b + G_{cd}w^c w^d\right)\\
	J &= \delta^a_c n_a w^c,
\end{aligned}
\end{equation}
Here the momentum and winding numbers $n$ and $w$ are $D$-component vectors of integers.

The average of such lattice theta functions on a Riemann surface of genus $g$ over the Narain moduli space has been studied by mathematicians, and admits a beautifully simple expression in terms of a non-holomorphic Eisenstein series of degree $g$. This remarkable relationship is known as the Siegel-Weil formula, the genus-one version of which can be stated as
\begin{equation}\label{eq:SiegelWeil}
	\langle Z(\tau)\rangle = {E_{D\over 2}(\tau)\over (\im\tau)^{D/2}|\eta(\tau)|^{2D}}.
\end{equation}
Here $\langle\cdot\rangle$ denotes the average over the Narain moduli space with respect to the natural homogeneous measure on ${\cal M}_D$, $Z(\tau)$ is the partition function of the CFT on a torus with modular parameter $\tau$, and $E_{D\over 2}$ is the real analytic Eisenstein series (features of which we review in detail in section \ref{sec:partitionFunction}). 

As described by \cite{Maloney:2020nni,Afkhami-Jeddi:2020ezh}, this formula suggests a new holographic duality relating the ensemble of Narain lattice CFTs with an exotic bulk theory that is \emph{perturbatively} equivalent to $U(1)^D\times U(1)^D$ Chern-Simons theory with action
\begin{equation}
	S_{\rm CS} = {1\over 2\pi}\sum_{j=1}^D\int\left(A_j\wedge dA_j - \widetilde A_j\wedge \widetilde d A_j\right).
\end{equation} 
We emphasize that this does not define the theory non-perturbatively, and that it must be supplemented with a particular instruction to sum over topologies.  The reason for this perturbative correspondence is the following. The real analytic Eisenstein series can be written in terms of a sum over orbits of the genus-one modular group $PSL(2,\ZZ)$
\begin{equation} 
	E_{s}(\tau) = \sum_{\gamma\in \Gamma_\infty\backslash PSL(2,\ZZ)}(\im\gamma\tau)^s,
\end{equation}
where $\Gamma_\infty$ is the subgroup of $PSL(2,\ZZ)$ that fixes $\im\tau$. Thus the Siegel-Weil formula (\ref{eq:SiegelWeil}) means that the averaged partition function can be written as a sum over modular images of the vacuum character of the $U(1)^D\times U(1)^D$ current algebra
\begin{equation}
	\langle Z(\tau)\rangle = \sum_{\gamma\in\Gamma_\infty\backslash PSL(2,\ZZ)}{1\over |\eta(\gamma\tau)|^{2D}}.
\end{equation}
In writing the Siegel-Weil formula like this we have used the fact that $(\im\tau)^{1/2}|\eta(\tau)|^2$ is modular invariant. 

The gravitational interpretation of the Siegel-Weil formula is facilitated by the fact that the averaged partition function can be written as a modular sum, following a similar logic as in the computation of the pure gravity partition function in \cite{Maloney:2007ud}. Indeed, the seed of the modular sum, the $U(1)^D\times U(1)^D$ vacuum character, is interpreted as the one-loop thermal partition function of $U(1)^D\times U(1)^D$ Chern-Simons theory on a three-manifold $D_2\times S^1$ whose boundary is a torus with modular parameter $\tau$.
This one-loop partition function is \cite{Giombi:2008vd,Porrati:2019knx,Maloney:2020nni}
\begin{equation}
	\mathcal{Z}_{D_2\times S^1}^{U(1)^{2D}}(\tau) = \left( ({\det '\Delta_0})^{3\over 2} \over (\det \Delta_1)^{\half} \right)^D = {1\over|\eta(\tau)|^{2D}},
\end{equation}
where $\det'\Delta_0$ is the (regularized) determinant of the scalar Laplacian on hyperbolic three-space $\mathbb{H}^3$ and $\Delta_1$ is the Laplacian for spin-one fields on $\mathbb{H}^3$. The result is a partition function which counts ``boundary photons," the edge modes associated with $U(1)^D\times U(1)^D$ Chern-Simons theory on $D_2\times S^1$. The modular sum in the genus-one Siegel-Weil formula (\ref{eq:SiegelWeil}) is then interpreted as a sum over topologies, and in particular as a sum over handlebody geometries with torus boundary which are labelled by $\Gamma_\infty\backslash PSL(2,\ZZ)$. This exotic gravitational theory was dubbed ``U(1) gravity'' in \cite{Afkhami-Jeddi:2020ezh}. We note that the sum is over all handlebodies with torus boundary, since every handlebody with torus boundary is obtained from $D_2\times S^1$ by a modular transformation. 

The geometry $D_2\times S^1$ is sometimes referred to as ``thermal $AdS_3$,'' since it can be obtained from Lorentzian $AdS_3$ by Wick-rotation and periodic identification of the Euclidean time coordinate. 
The other handlebodies with torus boundary have the same metric, and differ only by which coordinates are identified with the spatial circle and Euclidean time coordinate.  For example, the Euclidean BTZ black hole also has geometry $D_2 \times S^1$, except that now the Euclidean time cycle is contractible in the bulk rather than the spatial circle.  So Euclidean BTZ is also included in the sum over geometries encoded in the Siegel-Weil formula, where it is obtained from thermal $AdS_3$ by a modular $S$ transformation of the boundary modular parameter, $\tau\to -1/\tau$. This is the modular transformation which exchanges the space and time cycles of the boundary torus. Indeed, the entire family of $SL(2,\ZZ)$ black holes introduced in \cite{Maldacena:1998bw}, obtained from thermal $AdS_3$ by other $SL(2,\ZZ)$ modular transformations (where different linear combinations of the spatial and temporal cycles are taken to be contractible), are equal participants in the sum over geometries.

The bulk path integral as we have described it so far is very similar to the sum over geometries in pure $AdS_3$ quantum gravity as originally formulated in \cite{Maloney:2007ud}. However, in our bulk theory the boundary graviton is in a sense a composite of the boundary photon of the exotic theory, and indeed the statistics of the black hole microstates in the boundary theory (encoded by the statistics of primary operators in the averaged Narain CFT) are drastically different from those expected in pure gravity. For example, in the microcanonical ensemble at dimensions much larger than the central charge each member of the $SL(2,\ZZ)$ family of black holes gives a contribution to the density of states which scales like $\Delta^{c-2}$, unlike in pure gravity where the contribution of the Euclidean BTZ black hole dominates exponentially at high energy. Furthermore, in pure gravity the near-extremal spectrum has been argued to exhibit a square-root edge in the twist \cite{Maxfield:2019hdt,Maxfield:2020ale}, whereas in the Narain ensemble the near-extremal behaviour of the density of states is not quantitatively different from the high-energy growth, as we will see explicitly in (\ref{eq:narainDensityOfStates}). From the CFT point of view, these distinctions are rooted in the fact that in pure gravity, the vacuum is a ``censored'' state in the sense of \cite{Keller:2014xba}, and that there are null-state subtractions involved in the Virasoro vacuum character --- neither of which hold for the vacuum of the $U(1)^D\times U(1)^D$ chiral algebra.

Much of what was discussed above for the torus partition function can be generalized to partition functions on Riemann surfaces of arbitrary genus $g$.  The partition function of a Narain CFT on a genus $g$ surface is 
\be
Z_{\Sigma_g}(\Omega) = {\Theta(\Omega,m)\over (\det\im\Omega)^{{D\over 2}} |\det'\bar\partial_{\Sigma_g}|^D}
\ee
where $\Omega$ is the period matrix of the genus $g$ Riemann surface $\Sigma_g$ and $\det'\bar\partial_{\Sigma_g}$ is the determinant of the $\bar\partial$ operator on the Riemann surface (with zero-modes omitted).  This generalizes the factor of $\eta(\tau)$ in the genus-one case. 
The combination $(\det\im\Omega)^{{D\over 2}} |\det'\bar\partial_{\Sigma_g}|^D$ is modular-invariant on its own.
In computing the average over Narain moduli space, we must integrate $\Theta(\Omega,m)$ over the space of lattices ${\cal M}_D$.  This is given by a higher genus version of the
Siegel-Weil formula:
\begin{equation}\label{eq:higherGenusSiegelWeil}
	\langle Z_{\Sigma_g}(\Omega)\rangle = {E_{D\over 2}^{(g)}(\Omega)\over (\det\im\Omega)^{{D\over 2}} |\det'\bar\partial_{\Sigma_g}|^D}.
\end{equation}
where
\be
E_{s}^{(g)}(\Omega) \equiv \sum_{\gamma \in P \backslash Sp(2g,\Z)} (\det\im\gamma\Omega)^{s}
\ee
is an Eisenstein series of degree $g$.  Appendix \ref{app:eisensteinFourier} contains a review of various features of this Eisenstein series, including a definition of the subgroup $P$.  We refer to \cite{Maloney:2020nni} for more discussion of the derivation of equation (\ref{eq:higherGenusSiegelWeil}).

Much like the genus one case, the higher degree Eisenstein series can be written as a sum over images of the $U(1)^D\times U(1)^D$ genus-$g$ vacuum block under the genus $g$ modular group $Sp(2g,\ZZ)$, so that the averaged partition function can be interpreted as a sum over handlebody geometries with boundary $\Sigma_g$. Indeed, in \cite{Maloney:2020nni} it is shown that the seed of the modular sum can be interpreted as the partition function of boundary photons of $U(1)^D\times U(1)^D$ Chern-Simons theory on a particular handlebody; the denominator in equation (\ref{eq:higherGenusSiegelWeil}) is a conformal block for the $U(1)^D\times U(1)^D$ current algebra on the surface $\Sigma_g$.  However, there is to the best of our knowledge no a priori physical justification for the inclusion of \emph{only} handlebodies in the gravitational path integral; indeed, even in pure gravity with higher-genus boundary conditions, there are gravitational instantons that do not correspond to handlebodies (see e.g. \cite{Yin:2007at}). In our view, this feature is part of what makes the bulk theory so exotic -- but it is also what makes the bulk theory vastly simpler than pure Einstein gravity. Furthermore, the average (\ref{eq:higherGenusSiegelWeil}) only converges for $D>g+1$, so the bulk theory does not appear to be non-perturbatively well-defined; in particular for every value of the central charge there is a genus above which the averaged partition function diverges. It turns out that the higher-degree Eisenstein series that appears on the right-hand side of (\ref{eq:higherGenusSiegelWeil}) can be analytically continued in the complex $s$ plane outside of its naive range of convergence.  Although this can be used to obtain a finite expression in many cases,  it still turns out that for every value of $D$ there is at least one value of the genus for which this analytic continuation still diverges.\footnote{This is because the meromorphic continuation of the higher-degree Eisenstein series $E_s^{(g)}$ has poles at certain small quarter-integer values of $s$ \cite{Kalinin_1978}.}  The most conservative approach (which we will follow in this paper) is to only trust equation (\ref{eq:higherGenusSiegelWeil}) for genus $g<D-1$ which is not larger than the central charge.

In fact, as discussed in \cite{Maloney:2020nni}, one can make sense of the Siegel-Weil formula even when the boundary has multiple disconnected components.  This is particularly useful for diagnosing quantum chaos, and for studying the role of Euclidean wormholes in the path integral of the exotic bulk theory.  Indeed, the higher-genus generalization of the Siegel-Weil formula (\ref{eq:higherGenusSiegelWeil}) does not require $\Omega$ to be the period matrix of a Riemann surface; it only requires that the matrix $\Omega$ have positive imaginary part (i.e. that it is an element of the Siegel upper half-space $\mathcal{H}_n$). This includes the case where the Riemann surface is a disjoint union of Riemann surfaces $\Sigma_{g_i}$, in which case $\Omega$ is the direct sum of period matrices of the individual Riemann surfaces:
\begin{equation}
	\Omega = \bigoplus_{i=1}^n\Omega_i.
\end{equation}
This allows us to write the average of the product of partition functions in terms of a degree $g=\sum\limits_{i=1}^n g_i$ Eisenstein series
\begin{equation}\label{eq:multiBoundarySiegelWeil}
	\langle Z_{\Sigma_{g_1}}(\Omega_1)\cdots Z_{\Sigma_{g_n}}(\Omega_n)\rangle = {E^{(g)}_{D\over 2}(\Omega)\over \prod\limits_{i=1}^n (\det\im\Omega_i)^{D\over 2}|\det'\bar\partial_{\Sigma_{g_i}}|^D}.
\end{equation}
This average only converges for $\sum_{i=1}^n g_i < D - 1$ as noted above; we take this as perhaps an indication that sufficiently finely-grained observables are not captured by the simple ensemble-averaged theory considered in this paper. Importantly, this result for the averaged multi-point function does not factorize
\begin{equation}
	\langle Z_{\Sigma_{g_1}}(\Omega_1)\cdots Z_{\Sigma_{g_n}}(\Omega_n)\rangle \ne \prod_{i=1}^n\langle Z_{\Sigma_{g_i}}(\Omega_i)\rangle.
\end{equation}
This is a consequence of the fact that we are studying an ensemble of conformal field theories rather than any particular member of the ensemble. 

From the bulk point of view, we would like to interpret the sum (\ref{eq:multiBoundarySiegelWeil}) as the analog of the sum over Euclidean wormholes, i.e. the sum over bulk geometries which connect the different boundaries $\Sigma_{g_i}$.  
In particular, we note that the Eisenstein series is a sum over (a coset of) the modular group $Sp(2g,\Z)$, which acts on the cycles  $\bigoplus H^1(\Sigma_{g_i})$ of the boundary geometries in a way which respects the sympletic inner product between cycles.
Some of the terms in this sum are easy to interpret.  In particular, the terms where $\gamma$ is block diagonal correspond to bulk geometries which are just disconnected handlebodies; these give contributions to $\langle Z_{\Sigma_{g_1}}(\Omega_1)\cdots Z_{\Sigma_{g_n}}(\Omega_n)\rangle$ which  factorize.  The terms where $\gamma$ is not block diagonal are more interesting, as they are responsible for non-factorization.  For each such term in the sum, we interpret $\gamma$ as providing instructions for constructing a bulk configuration where certain boundary cycles are contractible in the bulk and others are not.   As we will see (and as was already discussed in \cite{Maloney:2020nni}) this includes bulk topologies which are simple Euclidean wormholes.  For example, in the two-boundary case this includes bulk configurations of the form $\Sigma_g \times I$.
We note, however, the the bulk interpretation of the sum is not as clear as in the handlebody case.  In particular, there does not appear to be a straightforward association between individual terms in the Eisenstein series and particular hyperbolic manifolds with a specified topology.\footnote{This is particularly clear in the case of torus boundaries, where one can prove that the only hyperbolic manifolds with torus boundaries are disconnected unions of handlebodies; these would correspond to the block diagonal terms in the Eisenstein series.}

Our primary focus will be the multi-point functions of the torus partition function, where
\begin{equation}
	\Omega = \diag(\tau_1,\ldots,\tau_n).
\end{equation}
Here the $\tau_i$ are the modular parameters of the individual tori, in which case the $n$-point function is computed by a degree-$n$ Eisenstein series. We will warm up by reviewing the $n=1$ case in the next section, before moving on to $n=2$ in section \ref{sec:twoPoint} and higher $n$ in section \ref{sec:higherPoint}.

\section{Warm-up exercise: the one-point function}\label{sec:partitionFunction}
We will begin by reviewing the Siegel-Weil formula in the case of a single genus-one boundary. In this case, the Siegel-Weil formula computes the averaged torus partition function. For each CFT in the ensemble, the torus partition function can be written as a thermal trace over the Hilbert space of the CFT on the circle
\begin{equation}\label{eq:partitionFunction}
	Z(\tau) = \Tr_{\mathcal{H}_{S^1}}(q^{L_0-{c\over 24}}\bar q^{\widetilde L_0-{c\over 24}}) = \Tr_{\mathcal{H}_{S^1}}(e^{-2\pi y(L_0+\widetilde L_0 - {c\over 12})+2\pi i x(L_0-\widetilde L_0)}).
\end{equation}
Here we have written $\tau = x+iy$, with $y$ proportional to the inverse temperature and $x$ an angular potential conjugate to the spin. We will be interested in the partition function that counts primary operators, which is related to the partition function (\ref{eq:partitionFunction}) by
\begin{equation}
	Z^p(\tau) = |\eta(\tau)|^{2D}Z(\tau).
\end{equation}

\subsection{Dissecting the Eisenstein series}
The Siegel-Weil formula for the one-point function of the genus-one partition function writes the averaged torus partition function in terms of a real-analytic Eisenstein series 
\begin{equation}\label{eq:genusOneEisenstein}
	\langle Z^p(\tau)\rangle = y^{-s}E_s(\tau) = \sum_{\gamma\in \Gamma_\infty\backslash PSL(2,\mathbb{Z})}{1\over |c\tau+d|^{2s}} = \sum_{\gamma\in \Gamma_\infty\backslash PSL(2,\mathbb{Z})} {1\over ( (cx+d)^2+c^2y^2)^{s}},
\end{equation}
where we have defined $s = {D\over 2}$ and\footnote{There is some degeneracy of notation between the central charge $c$ and the lower entry of a $PSL(2,\ZZ)$ element. The distinction should be clear from the context, however for this reason we will mostly use $D$ or $s$ as proxies for the central charge.}
\begin{equation}
	\gamma  = \mat{a}{b}{c}{d}\in PSL(2,\mathbb{Z}).
\end{equation}
The quotient by $\Gamma_\infty$ is by the space of upper triangular matrices in $PSL(2,\ZZ)$ that shift $x$ by an integer.  This quotient is necessary in order to render the sum (\ref{eq:genusOneEisenstein}) finite. The sum is then over a pair of coprime integers $(c,d)=1$. The sum (\ref{eq:genusOneEisenstein}) is convergent for $\re(s)>1$, but can be defined by analytic continuation away from a simple pole at $s=1$. The invariance of (\ref{eq:genusOneEisenstein}) by integer shifts of $x$ means that it admits the Fourier decomposition
\begin{equation}\label{eq:fourierDecomposition}
	\langle Z^p(\tau)\rangle = \sum_{j\in\ZZ}e^{2\pi i j x}\langle Z^p_j(y)\rangle,
\end{equation}
where $Z^p_j(y)$ counts primaries of spin $j$.

The Fourier decomposition of real analytic $SL(2,\ZZ)$ Eisenstein series is standard, but we will review it here to prepare for the more complicated higher-genus Eisenstein series that we will encounter later. It is convenient to separate out the contribution of the $c=0$ term, and for the remaining terms write $d = d' + nc$ where $d'\in \ZZ/c\ZZ$ such that $(c,d)=1$\footnote{We will refer to this subset as $(\ZZ/c\ZZ)^*$.} and $n$ is an integer. Then we have
\begin{equation}\label{eq:narainPoincare}
	\langle Z^p(\tau)\rangle = 1 + \sum_{c=1}^\infty\sum_{n\in\ZZ}\sum_{d'\in(\ZZ/c\ZZ)^*}c^{-2s}\left((x+n+d'/c)^2+y^2\right)^{-s}.
\end{equation}
We evaluate the sum over $n$ using the Poisson summation formula
\begin{equation}
	\langle Z^p(\tau)\rangle = 1 + \sum_{j\in\ZZ}\sum_{c=1}^\infty \sum_{d'\in (\ZZ/c\ZZ)^*}c^{-2s}e^{2\pi i j(x+d'/c)}\int_{-\infty}^\infty dn{e^{-2\pi i j n}\over (n^2+y^2)^s}.
\end{equation}
The sum over $j$ is precisely the Fourier decomposition (\ref{eq:fourierDecomposition}).

The remaining sum over $d'$ is a familiar object in number theory known as a Ramanujan sum
\begin{equation}
	c_c(j) = \sum_{d'\in(\ZZ/c\ZZ)^*}e^{2\pi i j d'/c},
\end{equation}
a special case of which gives the Euler totient function, $c_c(0) = \phi(c)$. The sum over $c$ is also computable for integer $j$, giving
\begin{equation}\label{eq:sumsOfRamanujanSums}
	\kappa_s(j)\equiv \sum_{c=1}^\infty c^{-2s}c_c(j) = \begin{cases} {\zeta(2s-1)\over \zeta(2s)},~ &j =0 \\ {\sigma_{2s-1}(j)\over |j|^{2s-1}\zeta(2s)},~&j\ne 0 \end{cases}.
\end{equation}

Putting the pieces together, we then have the following for the averaged partition function
\begin{equation}\label{eq:eisensteinFourier}
	\langle Z^p(\tau)\rangle = 1 + b_s(0)y^{1-2s} + \sum_{j=1}^\infty2\cos(2\pi j x)b_s(j)y^{\half-s}K_{s-\half}(2\pi j y),
\end{equation}
where
\begin{equation}\label{eq:beeEss}
b_s(j) = \begin{cases}{\sqrt{\pi}\Gamma(s-\half)\zeta(2s-1)\over \Gamma(s)\zeta(2s)},~& j=0 \\ {2\pi^s} \sigma_{2s-1}(j)\over j^{s-\half}\zeta(2s)\Gamma(s),~&j\ne 0 \end{cases}
\end{equation}
and $K$ is the modified Bessel function of the second kind.

As previously mentioned, although the Eisenstein series diverges for $\re(s)>1$, using the above expressions we can define it by analytic continuation away from this domain. An interesting example is the case $s=\half$ (i.e. $D=1$), in which case the corresponding Eisenstein series exactly vanishes
\begin{equation}
	\langle Z^p(\tau) \rangle_{D=1} = y^{-\half}E_{\half}(\tau) = 0,
\end{equation}
as can be seen directly by examining the appropriate limits of (\ref{eq:beeEss}). Of course, the integral over moduli space that defines the left-hand side diverges for $D=1$, so this should be regarded as a formal result associated with the regularization of the integral. Another interesting example is the case $D=2$, in which case the coefficient $b_1(0)$ diverges and the Eisenstein series cannot be rendered finite by analytic continuation (despite the fact that in this case the moduli space has finite volume).

\subsection{The density of states}
The density of primary states in the averaged theory can be extracted from our previous results by an inverse Laplace transform. 
In fact, the density of states can be derived more directly by summing over modular crossing kernels, which associate a particular contribution to the density of states to each term in the sum over handlebodies; 
this alternate derivation is given in appendix \ref{app:genusOneCrossingKernels}. Although any particular Narain lattice CFT in the ensemble has a discrete spectrum of primary states, the effect of the ensemble average is to render the density of states continuous, with support above the unitarity bound in each spin sector.  

First we note that at low temperature ($y\to\infty$) we have
\begin{equation}
	\langle Z^p_j(y)\rangle \to \delta_{j,0},
\end{equation}
reflecting the fact that the only operator with dimension zero is the identity operator. Meanwhile, at high temperature ($y\to 0$), we have
\begin{equation}
	\langle Z^p_j(y)\rangle \sim \begin{cases} b_s(0)y^{1-2s},~& j=0 \\ \half (\pi |j|)^{\half-s}\Gamma(s-1/2)b_s(j)y^{1-2s},~&j\ne 0\end{cases}.
\end{equation}
The polynomial growth reflects sub-Cardy growth of the density of states at large dimension due to the fact that $c= c_{\rm currents}$, which we will see very explicitly momentarily.

The density of states computed from the inverse Laplace transform of (\ref{eq:eisensteinFourier})
\begin{equation}
	\langle\rho^p_j(\Delta)\rangle = 2\pi \int_{\mathcal{C}}{dy\over 2\pi i}\,\langle Z^p_j(y)\rangle e^{2\pi \Delta y},
\end{equation}
where $\mathcal{C}$ is a suitable vertical contour in the $y$ plane, yields:
\begin{equation}\label{eq:narainDensityOfStates}
	\langle \rho^p_j(\Delta)\rangle = \begin{cases} \delta(\Delta) + {2\pi^{2s}\zeta(2s-1)\over \zeta(2s)\Gamma(s)^2}\Delta^{2s-2},~ &j=0 \\ {2\pi^{2s}\sigma_{2s-1}(j)\over |j|^{2s-1}\Gamma(s)^2\zeta(2s)}(\Delta^2-j^2)^{s-1},~ & j\ne 0\end{cases}.
\end{equation}
This matches precisely the result derived in \cite{Afkhami-Jeddi:2020ezh} (using other methods) and more directly in appendix \ref{app:genusOneCrossingKernels} using crossing kernels. We emphasize that in each spin sector the density of states is supported everywhere above the unitarity bound, $\Delta\ge |j|$. The delta function at $\Delta=0$ in the scalar sector represents the identity operator. These exact results very explicitly demonstrate the sub-Cardy growth of the averaged density of states at large dimension. Unlike the corresponding density of states in pure gravity \cite{Maloney:2007ud,Keller:2014xba,Benjamin:2019stq} the density of states obtained from averaging Narain lattice CFTs is manifestly positive everywhere.  That the density of states is continuous is a direct consequence of the ensemble average.

\subsection{The semiclassical limit and comparison to pure 3D gravity}

The density of states (\ref{eq:narainDensityOfStates}) exhibits some interesting features in the large $s$ limit, which is the semiclassical limit. Although the density of states is supported above the unitarity bound $\Delta\ge |j|$, at large $s$ we have\footnote{Here we have temporarily defined $\tau\equiv \Delta - |j|$ to be the twist, not to be confused with the modular parameter of the boundary torus.} 
\begin{equation}
	\langle\rho^p_j(\tau)\rangle = {s\over \pi \tau(\tau+2|j|)}\left({\pi e\over s}\sqrt{\tau(\tau+2|j|)}\right)^{2s}\left(1+\mathcal{O}(s^{-1})\right).
\end{equation}
As emphasized in \cite{Afkhami-Jeddi:2020ezh}, at fixed spin $j$ this is exponentially suppressed until the twist has attained the bound
\begin{equation}
\tau \gtrsim {s\over \pi e}.
\end{equation}
Furthermore, at large $s$ the averaged partition function is dominated by the contribution from the $c=1$ instantons in (\ref{eq:narainPoincare}), up to terms that are non-perturbatively suppressed in $s$. This is because the sums of Ramanujan sums (\ref{eq:sumsOfRamanujanSums}) become exponentially close to $1$ in the large-$s$ limit. So although the density of states is supported everywhere above the unitarity bound, there is a sense in which it is negligible until the twist is of order the central charge in the semiclassical limit, which is a signature of a holographic system.

As in the case of the path integral of pure three-dimensional gravity with negative cosmological constant, we interpret the Poincar\'e series that appears in the torus partition function (\ref{eq:genusOneEisenstein}) as a sum over the $PSL(2,\ZZ)$ black holes, which are smooth saddle points of the gravitational path integral. Indeed, in the large $s$ limit at fixed inverse temperature $y = {\beta\over 2\pi}$, the partition function (\ref{eq:genusOneEisenstein}) exhibits a Hawking-Page phase transition
\begin{equation}
	Z^p(\tau) \approx 1 + \left(2\pi\over \beta\right)^{2s}.
\end{equation}
For $\beta > 2\pi$, the first term, representing the contribution of thermal AdS$_3$, dominates. There is a first-order phase transition as $\beta$ is decreased below $2\pi$ and the Euclidean BTZ black hole becomes the dominant saddle. Although the technical details are different, this mirrors what happens in pure three-dimensional gravity.  

It is important to note, however, that the contributions from the full family of  $PSL(2,\ZZ)$ black holes are, in a sense, more important in $U(1)$ gravity than in pure gravity.  The  $PSL(2,\ZZ)$ black holes which are not Euclidean BTZ correspond to terms in (\ref{eq:genusOneEisenstein}) with $c>1$.  These give contributions to the partition function $Z^p(\tau)$ which are also of order $\left(2\pi\over \beta\right)^{2s}$ but with a coefficient which is exponentially suppressed relative to the $c=1$ contribution at large $s$.  In other words, they give contributions to the entropy which are subleading (relative to the BTZ contribution) at large central charge, but will not be subleading at high temperature and fixed central charge.  This should be contrasted with the pure gravity case, where these contributions are subleading both at large central charge and at high temperature with fixed central charge. Similar statements hold in the microcanonical ensemble, where each $PSL(2,\ZZ)$ black hole gives a contribution to the density of states which scales like $\Delta^{2s-2}$ at large $\Delta$, but with the a coefficient that is (at large central charge) dominated by the contribution from Euclidean BTZ. One simple way to see this is to inspect the individual contributions to the density of states using the crossing kernel approach in appendix \ref{app:genusOneCrossingKernels}.

\section{The two-point function}\label{sec:twoPoint}

We now consider the Siegel-Weil formula which computes the correlation function of two primary-counting partition functions in terms of a degree-two Eisenstein series:\footnote{In this section and whenever discussing the modular group for genus greater than one, $D$ will exclusively refer to the lower constituent of an $Sp(2g,\mathbb{Z})$ element, while $s = {D\over 2}$ will be used as a proxy for the central charge.}
\begin{equation}\label{eq:twoPointFunction}
	\langle Z^p(\tau_1)Z^p(\tau_2)\rangle =  {E^{(2)}_s(\Omega)\over (y_1y_2)^s} = \sum_{(C,D) = 1}|\det(C\Omega+D)|^{-2s}.
\end{equation}
Here
\begin{equation}
	\Omega = \mat{\tau_1}{0}{0}{\tau_2}
\end{equation}
is a diagonal element of the Siegel upper half-space $\mathcal{H}_2$, with $\tau_i = x_i+iy_i$ the modular parameters of the two tori.  On the right hand side of equation (\ref{eq:twoPointFunction}) we have written the Eisenstein series as a sum over $2\times2$ matrices $C$ and $D$.  The ``co-prime" condition, written $(C,D)=1$, is the statement that these matrices  make up the lower row of an $Sp(4,\Z)$ matrix, which we review in appendix \ref{app:eisensteinFourier}.
The goal of this section is to compute (\ref{eq:twoPointFunction}) explicitly.  Before doing so, however, we will comment first on the interpretation of this equation as a sum over topologies.

We note that equation (\ref{eq:twoPointFunction}) can be obtained from the averaged genus-two partition function in the limit where the off-diagonal element of the period matrix vanishes. This is the limit in which a genus-two Riemann surface degenerates into two tori glued together at a point. We note that in pure AdS$_3$ gravity, this limit of the genus-two partition function is singular.  This is because the ground state of a CFT on a circle has negative energy, so the amplitude for the identity operator to propagate 
down the narrow tube separating the tori will diverge in the limit that the tube is becoming very long. In the CFT language, this is the statement that the genus 2 partition function diverges like $q^{-c/12}$ in this degeneration limit, where $q$ is related to the modular parameter of the long tube.  A remarkable feature of the Narain ensemble is that, once we divide out by the appropriate conformal blocks to compute the primary-counting partition function, this limit is finite.  This is a direct consequence of the fact that we are considering theories with $c=c_{\rm currents}$.

This picture has an appealing bulk interpretation as well. As explained in \cite{Maloney:2020nni}, the modular sum appearing in (\ref{eq:multiBoundarySiegelWeil}) is best understood as a sum over indecomposable Lagrangian sublattices. For a Riemann surface $\Sigma_g$ of genus $g$, the first homology group $H_1(\Sigma_g;\mathbb{Z})$ is a lattice generated by the $A$- and $B$-cycles of the Riemann surface. A Lagrangian sublattice $\Gamma\subset H_1$ is a primitive sublattice of rank $g$ for which the intersection pairings all vanish. A simple example of a Lagrangian sublattice is defined by the homology classes of the set of $A$-cycles of the Riemann surface. In this way, each term in the modular sum in (\ref{eq:twoPointFunction}) can be regarded as providing a set of instructions for identifying the contractible cycles in a corresponding bulk topology: it is the one where the sublattice $\Gamma$ is contractible. It is important to note that this procedure is somewhat ambiguous, however, as a choice of cycles to be made contractible does not uniquely specify a bulk topology.  When the boundary is connected, a choice of contractible cycles is sufficient to uniquely identify a bulk handlebody -- that is to say, a three-manifold whose topology is defined by embedding our surface in $\RR^3$ and taking the interior of our surface in $\RR^3$.  But there are many other topological 3-manifolds (non-handlebodies) for which the same configuration of cycles will be contractible in the bulk. We interpret this as the the statement that $U(1)$ gravity, at least as defined by the Siegel-Weil formula, is only sensitive to some (but not all) of the data of the bulk topology. The same phenomenon will occur when we take a pinching limit of a genus 2 surface to get a union of two disconnected tori.  In particular, when we take the pinching limit the bulk topologies we obtain are not uniquely specified by a choice of Lagrangian sublattice.  Instead, we should regard the topologies so obtained as representatives of an entire (infinite) class of topological wormholes associated with a particular sublattice.  Nevertheless, we will see that in many cases it is easy to obtain representative wormhole topologies that have a simple bulk interpretation.

\begin{figure}[h]
  \centering
  \includegraphics[width=.99\textwidth]{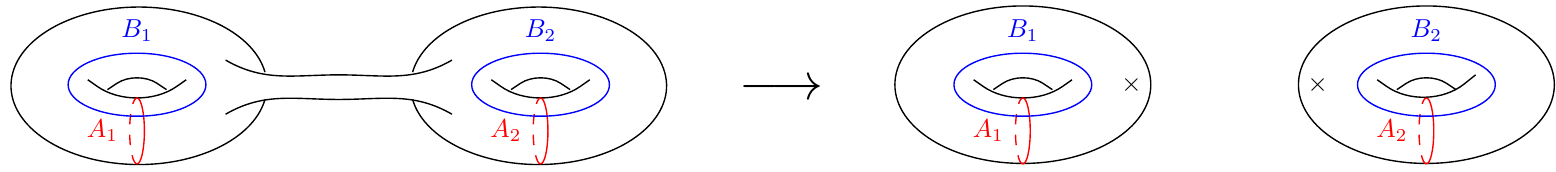}
  \caption{The degeneration of a genus-two handlebody into two solid tori. The $A$-cycles of the tori are contractible in the bulk, corresponding to the manifold $M_{0,\id}$.}\label{fig:degeneratingHandlebody}
\end{figure}

We will now explain how this works for three simple examples of terms in the modular sum. To start, consider a handlebody (i.e. a solid genus 2 surface) where one fills some genus-two boundary.  Denote by $A_1,A_2$ the cycles of the boundary which are contractible in the handlebody, and by $B_1,B_2$ the non-contractible cycles, with intersection numbers $\langle A_i, B_j\rangle = \delta_{ij}$. If one pinches the cycle $A_1\,B_1\,A_1^{-1}\,B_1^{-1}$ (or equivalently $A_2\,B_2\,A_2^{-1}\,B_2^{-1}$), then one obtains two solid tori (see figure \ref{fig:degeneratingHandlebody}). If we take the contractible cycles of each boundary torus to be the spatial cycles, then this is simply $\text{TAdS}_3\cup \text{TAdS}_3$, where $\text{TAdS}_3$ is thermal AdS$_3$. We will refer to this three-manifold as $M_{0,\id}$.

To understand how this works for other terms in the modular sum, one must use the fact that for each element $\gamma \in Sp(4,\mathbb{Z})$ with lower entries $C,D$, the corresponding lattice $\Gamma$ is\footnote{Here we are following the notation of \cite{Maloney:2020nni}: $H_1(\Sigma_2;\mathbb{Z})$ is identified with the integer lattice in 4 dimensions spanned by vectors $(b_1, b_2, a_1, a_2)$ where the integers $b_i$ and $a_i$ denote the number of times a cycle wraps $B_i$ and $A_i$, respectively.  So equation (\ref{eq:contractibleCycles}) defines a rank two sublattice sitting inside the integer lattice in four dimensions.}
\begin{equation}\label{eq:contractibleCycles}
  (0,\mathbf{a})\gamma = \left(\mathbf{a}\cdot C,\, \mathbf{a}\cdot D\right).
\end{equation}
This is to be regarded as a sublattice of $H_1(\Sigma_2;\mathbb{Z})$ that determines which cycles are contractible in the bulk.
In this way we see that $C$ and $D$ permute these contractible cycles, so that the sum over $(C,D)$ is indeed a kind of sum over wormholes.  
The contribution  $M_{0,\id}$ (i.e. $\text{TAdS}_3\cup \text{TAdS}_3$) is just the simplest example.
As another simple example, consider the term in (\ref{eq:twoPointFunction}) with $C=\id,\, D=0$, in which case according to (\ref{eq:contractibleCycles}) 
the contractible cycles are now:
\begin{equation}
\begin{aligned}
  M_{\id,0}:~~\widetilde A_1 &= B_1\\ 
  \widetilde A_2 &= B_2.
\end{aligned}
\end{equation}
In other words, the $B$-cycles of the surface are now taken to be contractible.
One can then start with the handlebody where these $B$ cycles are contractible and then pinch the cycle that is the $\gamma$ image of $A_1\,B_1\,A_1^{-1}\,B_1^{-1}$.  This leads to a disconnected three-manifold that where the time cycles of the tori are now contractible. We can therefore identify $M_{\id, 0}$ with the three-manifold that is simply the union of two Euclidean BTZ black holes. 

The other terms in (\ref{eq:twoPointFunction}) are handled similarly: one starts with a connected 3-manifold (such as a handlebody) in which the cycles given by (\ref{eq:contractibleCycles}) are contractible in the bulk. One then pinches the cycle that is the $\gamma$ image of $A_1\,B_1\,A_1^{-1}\,B_1^{-1}$ to obtain a bulk topology which can be associated with the $(C,D)$ term in the sum; we refer to this contribution as $M_{C,D}$. The modular sum (\ref{eq:twoPointFunction}) is then interpreted as a sum over these contributions.  When $C$ and $D$ are diagonal, these correspond to disconnected topologies, and the sum reproduces the disconnected sum over $SL(2,\mathbb{Z})$ black holes to reconstruct $\langle Z^p(\tau_1)\rangle\langle Z^p(\tau_2)\rangle$. When $C$ and/or $D$ are off-diagonal, the corresponding topologies are connected. We will see that some of them correspond to simple topologies equivalent to $T^2\times I$ where $I$ is an interval (see the discussion in section \ref{subsec:torusWormhole}), but the others will correspond to more topologically complicated three-manifolds.

\begin{figure}[h]
  \centering
  \includegraphics[width=.45\textwidth]{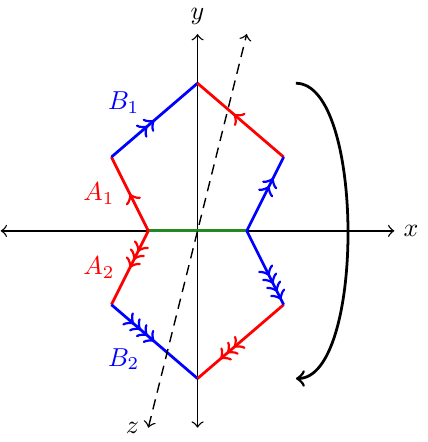}
  \caption{Here we illustrate how to construct the manifold $M_{C_0,D_0}$ from the pinching limit of a quotient of hyperbolic three-space. The statement that the cycles given by (\ref{eq:torusWormholeContractibleCycles}) are contractible in the bulk can be implemented (up to an orientation reversal that flips the sign of one of the $B$ cycles) by ``filling in'' the three-manifold by rotating the top half through the bulk around the $x$-axis by an angle of $\pi$ as indicated by the bold arrow so that $A_1$ is identified with $A_2$ and $B_1$ is identified with $B_2$. The manifold $M_{C_0,D_0}$ is then obtained by pinching the cycle $A_1\,B_1\,A_1^{-1}\,B_1^{-1}$, which is labelled in green. The resulting topology is that of a torus times an interval. One can even endow this with a constant negative curvature metric by replacing the straight line segments with circles that are identified via Mobius transformations which become hyperbolic isometries in the bulk.  Strictly speaking, this gives a constant negative curvature metric whose boundary is two copies of a punctured torus, rather than just two copies of the torus; this had to be the case, of course, since there is no smooth constant negative curvature manifold whose conformal boundary is the disconnected union of two tori.}\label{fig:quotientWormhole}
\end{figure}

As a concrete example of a term in the modular sum associated with a simple connected topology, consider the case where\footnote{This example was also discussed in \cite{Maloney:2020nni}.}
\begin{equation}\label{eq:torusWormholeContractibleCycles}
\begin{aligned}
  C &= C_0 = \begin{pmatrix} 1 & 1 \\ 0 & 0 \end{pmatrix}\\
  D &= D_0 = \begin{pmatrix} 0 & 0 \\ -1 & 1 \end{pmatrix}.
\end{aligned}
\end{equation}
According to (\ref{eq:contractibleCycles}), the contractible cycles in the bulk are now given by
\begin{equation}
\begin{aligned}
  M_{C_0,D_0}:~~\widetilde A_1 &= B_1 + B_2 \\
  \widetilde A_2 &= - A_1 + A_2.
\end{aligned}
\end{equation}
A natural bulk topology with these particular cycles being contractible is just the Euclidean wormhole $T^2\times I$. As explained in \cite{Maloney:2020nni}, the relative sign difference between the identifications in ${\widetilde A}_1$ and ${\widetilde A}_2$ arises from an orientation reversal of one of the boundaries relative to the other. In fact, it is even possible to obtain this Euclidean wormhole by taking a pinching limit of a hyperbolic three manifold with genus two boundary, as indicated in figure \ref{fig:quotientWormhole}.

We now turn to the explicit computation of $\langle Z^p(\tau_1)Z^p(\tau_2)\rangle$ given by equation (\ref{eq:twoPointFunction}).  The main technical challenge is to obtain a complete set of representatives of the pair of ``coprime'' matrices $(C,D)$. This is described beautifully in \cite{Maa__1971}, whose results we now briefly review (see appendix \ref{app:eisensteinFourier} for more details). The complete set of representatives of such matrices divides into three cases depending on the rank of the matrix $C$. So we will write
\begin{equation}
	\langle Z^p(\tau_1)Z^p(\tau_2)\rangle = 1 + \langle Z^p(\tau_1)Z^p(\tau_2)\rangle_1 + \langle Z^p(\tau_1)Z^p(\tau_2)\rangle_2,
\end{equation}
where the subscript denotes the contributions from terms where $C$ has the specified rank.  

\subsection{Rank 1}\label{subsec:rankOne}
We will first consider the case in which the matrix $C$ has rank one. We will simply state the results, referring to appendix \ref{app:eisensteinFourier} and \cite{Maa__1971} for details.  A complete set of representatives when $C$ has rank one is labelled by two pairs of coprime integers: $(c,d)=1$ with $c\ne 0$, and $(m,n)=1$. Since $m,n$ are coprime we can find integers $p,q$ such that $U = \mat{m}{p}{n}{q}$ is unimodular, and define $C$ and $D$ via
\begin{equation}
	C = \mat{c}{0}{0}{0} U^T,~~~~~D = \mat{d}{0}{0}{1} U^{-1}.
\end{equation}
Then
\begin{equation}
	\det(C\Omega+D) = c(m^2\tau_1+n^2\tau_2)+d \equiv c \tau_{m,n}+d,
\end{equation}
where
\begin{equation}
	\tau_{m,n} = m^2\tau_1+n^2\tau_2 \equiv x_{m,n}+iy_{m,n}.
\end{equation}
The sum over $(c,d)$ is precisely that involved with the genus-one Eisenstein series with a modified value of the modular parameter, $\tau = \tau_{m,n}$
\begin{equation}\label{eq:twoPointFunctionRankOne}
	\langle Z^p(\tau_1) Z^p(\tau_2)\rangle_1 = \sum_{(m,n)=1}\sum_{\substack{(c,d)=1\\c\ne 0}}{1\over |c\tau_{m,n}+d|^{2s}} = \sum_{(m,n)=1}\left(y_{m,n}^{-s}E_s(\tau_{m,n})-1\right) = \sum_{(m,n)=1}\left(\langle Z^p(\tau_{m,n})\rangle-1\right).
\end{equation}
We can then use the method of the previous section to write this as
\begin{equation}\label{eq:twoPointRank1}
	\langle Z^p(\tau_1) Z^p(\tau_2)\rangle_1 = \sum_{(m,n)=1}\left(b_s(0)y_{m,n}^{1-2s}+\sum_{j=1}^\infty2\cos(2\pi j x_{m,n})b_s(j)y_{m,n}^{\half-s}K_{s-\half}(2\pi j y_{m,n})\right).
\end{equation}
With this rewriting it is straightforward to decompose into sectors of definite spin with respect to either boundary via a Fourier transform. For instance, the correlator of spinless primaries is given by
\begin{equation}
	\langle Z^p_0(y_1)Z^p_0(y_2)\rangle_1 = b_s(0)\sum_{(m,n)=1}(m^2 y_1+n^2y_2)^{1-2s}.
\end{equation}
In the case that just one of the angular momenta vanishes, the correlator reduces to the single-boundary correlator
\begin{equation}
	\langle Z^p_0(y_1)Z^p_j(y_2)\rangle_1 = \langle Z^p_j(y_2)\rangle,~j\ne 0.
\end{equation}
This indicates that the terms in the $Sp(4,\ZZ)$ sum in which $C$ has rank one couple spinless and spinning states only through purely disconnected contributions to the two-point function. 

In the case that both angular momenta are non-vanishing, we have
\begin{equation}
	\langle Z^p_{j_1}(y_1)Z^p_{j_2}(y_2)\rangle_1 = \sum_{j\ne 0} \sum_{\substack{(m,n)=1\\ m^2 j = j_1,n^2 j = j_2}}b_s(j)y_{m,n}^{\half-s} K_{s-\half}(2\pi j y_{m,n}).
\end{equation}
In particular, when the spins are equal, only the $(m,n) = (1,1)$ term survives and the rank-one contribution to the two-point function is equal to a one-point function with a modular parameter that is the sum of the individual modular parameters
\begin{equation}
	\langle Z^p_{j}(y_1) Z^p_j(y_2)\rangle_1 = \langle Z^p_j(y_1+y_2)\rangle.
\end{equation}
When we consider the spectral form factor later on, this is precisely the term that will lead to the plateau. This (and indeed all terms in (\ref{eq:twoPointRank1}) with $(m,n)\ne (1,0)$ or $(0,1)$) is an example of a genuinely connected contribution to the two-point function, reflecting the fact that the bulk theory does not factorize.
Exactly such a term was described in section 3.2 of \cite{Maloney:2020nni}, where it was associated with a bulk wormhole with topology $T^2\times I$.

\subsection{Rank 2}\label{subsec:rankTwo}
When $C$ has rank two we can define $P = C^{-1}D$, which is a symmetric matrix with rational entries. The set of such matrices is precisely the complete set of representatives of coprime matrices $(C,D)$ with $\rank C =2$ \cite{Maa__1971}. The rank-two contribution to the two-point function is then given by the following sum
\begin{equation}
  \langle Z^p(\tau_1) Z^p(\tau_2) \rangle_2 = \sum_{\substack{P = P^T \\ \text{rational}}} \nu(P)^{-2s}|\det(\Omega+P)|^{-2s}.
\end{equation} 
The prefactor $\nu(P)$, which is the determinant of $C$, is defined in terms of $P$ as follows. There exist unimodular matrices $U,V$ such that
\begin{equation}
  P = U \mat{p_1\over q_1}{0}{0}{p_2\over q_2}V^{-1},
\end{equation}
where $p_i$ and $q_i$ are coprime integers with $q_i>0$. The ratios $p_i/q_i$ are known as the elementary divisors of $P$, and $\nu(P)$ is defined to be the product of the denominators of these elementary divisors: $\nu(P) = q_1q_2$. In what follows it will be important that $\nu(P)$ is invariant under shifts of $P$ by an integral matrix.

To digest this sum it is convenient to further decompose $P = R+N$, where $R$ is a symmetric matrix with rational entries between zero and one, and $N$ is a symmetric integral matrix. Then we have
\begin{equation}
	\det(\Omega+R+N) = (n_1+r_1+\tau_1)(n_2+r_2+\tau_2)-(n+r)^2,~N = \mat{n_1}{n}{n}{n_2},~R = \mat{r_1}{r}{r}{r_2}.
\end{equation}
We can then write the rank-two contribution to the two-point function in terms of sums over three integers and three rational numbers between zero and one
\begin{equation}\label{eq:eisensteinRankTwo}
	\langle Z^p(\tau_1)Z^p(\tau_2)\rangle_2 = \sum_{n_1,n_2,n\in\ZZ}\sum_{r_1,r_2,r\in\mathbb{Q}/\ZZ}\nu(R)^{-2s}(\widetilde n^4+2\widetilde n^2(y_1y_2-\widetilde n_1\widetilde n_2)+(\widetilde n_1^2+y_1^2)(\widetilde n_2^2+y_2^2))^{-s},
\end{equation}
where
\begin{equation}
\begin{aligned}\label{eq:definedTildedNs}
	\widetilde n_i &= n_i + r_i + x_i\\
	\widetilde n &= n+r.
\end{aligned}
\end{equation}
Note that that the $n=r=0$ term in (\ref{eq:eisensteinRankTwo}) corresponds to part of the disconnected contribution to the two-point correlator, so that
\begin{equation}\label{eq:rankTwoDisconnected}
	\langle Z^p(\tau_1)Z^p(\tau_2)\rangle_2 \supset \left(\langle Z^p(\tau_1)\rangle-1\right)\left(\langle Z^p(\tau_2)\rangle -1\right).
\end{equation}
All other terms contribute to the connected two-point function.

We can use similar techniques as in the one-boundary and rank-one contributions to decompose this result into sectors of definite spin, in particular applying the Poisson summation formula to the sums over $n_i$ to get
\begin{equation}
	\langle Z^p_{j_1}(y_1)Z^p_{j_2}(y_2)\rangle_2 = \sum_{\substack{r_1,r_2\in\mathbb{Q}/\mathbb{Z} \\ r \in \mathbb{Q}}}\nu(R)^{-2s}e^{2\pi i (j_1r_1+j_2r_2)}\int_{-\infty}^\infty dn_1 dn_2 \, {e^{-2\pi i(j_1 n_1+j_2 n_2)}\over (r^4 + 2r^2(y_1y_2-n_1n_2)+(n_1^2+y_1^2)(n_2^2+y_2^2))^s}.
\end{equation}

Following the discussion of the Fourier decomposition of the higher-degree Eisenstein series in appendix \ref{app:eisensteinFourier}, this can be further rewritten in terms of more physical variables as
\begin{equation}
\begin{aligned}\label{eq:twoPointFunctionRankTwoFixedSpin}
	&\langle Z^p_{j_1}(y_1)Z^p_{j_2}(y_2)\rangle_2\\ 
	=& {4\pi^{4s}\over \Gamma_2(s)^2}\sum_{j\in \half\mathbb{Z}}S_{s,2}(J)\int_{\mathcal{P}}d\Delta_1 d\Delta_2 d\Delta\, e^{-2\pi(\Delta_1 y_1 +\Delta_2 y_2)} \left[\prod_{\epsilon= \pm}\left((\Delta_1-\epsilon j_1)(\Delta_2-\epsilon j_2)-(\Delta-\epsilon j)^2\right)\right]^{s-{3\over 2}},
\end{aligned}
\end{equation}
where we have defined the matrix
\begin{equation}
	J = \mat{j_1}{j}{j}{j_2},
\end{equation}
$S_{s,2}(J)$ is Siegel's ``singular series'' defined in (\ref{eq:SiegelSingularSeries}), and the integration region $\mathcal{P}$ is defined as the domain in which the matrices $\mat{\Delta_1\mp j_1}{\Delta \mp j}{\Delta \mp j}{\Delta_2 \mp j_2}$ are positive-definite.

\subsection{Density-density correlations}
We can use the results from the previous subsections to compute the two-point correlation function of the density of primary states by double inverse Laplace transform of the two-point function (\ref{eq:twoPointFunction})
\begin{equation}
	\langle \rho^p_{j_1}(\Delta_1)\rho^p_{j_2}(\Delta_2)\rangle = (2\pi)^2\int_{\mathcal{C}_1}{dy_1\over 2\pi i}{\int_{\mathcal{C}_2}}{dy_2\over 2\pi i}\,\langle Z^p_{j_1}(y_1)Z^p_{j_2}(y_2)\rangle e^{2\pi(y_1\Delta_1+y_2\Delta_2)},
\end{equation}
where $\mathcal{C}_i$ are suitable vertical contours in the $y_i$ planes.

It is convenient to separately study the rank-one and rank-two contributions to the density-density correlations. The rank-one contributions (\ref{eq:twoPointFunctionRankOne}) can be written in terms of an infinite sum over one-point functions with a modified temperature, and so lead to a sum of delta function-localized contributions to the two-point function of the density of states. The rank-one contributions are dominant over the rank-two contributions at low temperatures $y_1,y_2\gg 1$, and so we conclude that the density-density pair correlations are localized near threshold $\Delta_1,\Delta_2 \gtrsim 0$. For non-vanishing spins $j_1,j_2$, the rank-one contributions to the two-point function of the density of states reads
\begin{equation}
	\langle \rho^p_{j_1}(\Delta_1)\rho^p_{j_2}(\Delta_2)\rangle_1 = \sum_{j\ne 0} \sum_{\substack{(m,n)=1\\ m^2 j = j_1,n^2 j = j_2}}{2\pi^{2s}\sigma_{2s-1}(j)\over |j|^{2s-1}\zeta(2s)\Gamma(s)^2}\left[\left({\Delta_1\over 2m^2}+{\Delta_2\over 2n^2}\right)^2-j^2\right]^{s-1}\delta(n^2\Delta_1-m^2\Delta_2).
\end{equation}
In particular, in the case that the spins are equal, we have
\begin{equation}
	\langle \rho^p_{j}(\Delta_1)\rho^p_{j}(\Delta_2)\rangle_1 = {2\pi^{2s}\sigma_{2s-1}(j)\over |j|^{2s-1}\zeta(2s)\Gamma(s)^2}\left[\left({\Delta_1+\Delta_2\over 2}\right)^2-j^2\right]^{s-1}\delta(\Delta_1-\Delta_2).
\end{equation}
Finally, the rank-one contributions to the density-density correlations in the scalar sector are given by
\begin{equation}
\begin{aligned}
	\langle \rho^p_0(\Delta_1)\rho^p_0(\Delta_2)\rangle_1 =& \, {2\pi^{2s}\zeta(2s-1)\over \zeta(2s)\Gamma(s)^2} \bigg[\Delta_1^{2s-2}\delta(\Delta_2)+\Delta_2^{2s-2}\delta(\Delta_1)\\
	& \, +\sum_{\substack{(m,n)=1 \\ m,n >0}}\left({\Delta_1\over 2m^2}+{\Delta_2\over 2n^2}\right)^{2s-2}\delta(n^2\Delta_1-m^2\Delta_2)\bigg].
\end{aligned}
\end{equation}

The rank-two contributions to the density-density correlations can be read off from (\ref{eq:twoPointFunctionRankTwoFixedSpin}) as
\begin{equation}\label{eq:rankTwoDensityDensity}
	\langle \rho^p_{j_1}(\Delta_1)\rho^p_{j_2}(\Delta_2)\rangle_2 = {4\pi^{4s}\over \Gamma_2(s)^2}\sum_{j\in\half\mathbb{Z}}S_{s,2}(J)\int_{\mathcal{P}_{\Delta_i,j_i,j}}d\Delta\, \prod_{\epsilon=\pm}\left( (\Delta_1-\epsilon j_1)(\Delta_2-\epsilon j_2)-(\Delta-\epsilon j)^2\right)^{s-{3\over 2}},
\end{equation}
where the integration region $\mathcal{P}_{\Delta_i,j_i,j}$ is defined for fixed $\Delta_i,j_i,j$ as the region for which the matrices $\mat{\Delta_1\pm j_1}{\Delta\pm j}{\Delta\pm j}{\Delta_2\pm j_2}$ are positive-definite. However, this expression is not particularly illuminating as the integral over the funny domain in $\Delta$ obscures the dependence on the dimensions $\Delta_1,\Delta_2$. For the purposes of computing the two-point function of the density of states, it turns out to be \emph{much} simpler to work in an ensemble of fixed $U(1)^D\times U(1)^D$ charges, as shown in \cite{Datta:2021ftn}. There, the rank-two contribution to the two-point function of the density of states in an ensemble of fixed $U(1)$ charges $Q_i^I,\bar Q_i^I$ (where $i=1,2$ and $I=1,\ldots, D$) was found to be
\begin{equation}
\begin{aligned}\label{eq:fixedChargeTwoPoint}
	\langle \rho^p_{j_1}(\Delta_1,Q^I_1,\bar Q^I_1)\rho^p_{j_2}(\Delta_2,Q^I_2,\bar Q^I_2)\rangle_2 =& \, \widetilde S_{s,2}(Q)\delta\left(\Delta_1-\half(Q_1^2+\bar Q_2^2)\right)\delta\left(\Delta_2-\half(Q_2^2+\bar Q_2^2)\right) \\
	& \, \times \delta\left(j_1-\half(Q_1^2-\bar Q_1^2)\right)\delta\left(j_2-\half (Q_2^2-\bar Q_2^2)\right).
\end{aligned}
\end{equation} 
Here, $\widetilde S_{s,2}$ is a generalization of Siegel's singular series
\begin{equation}
	\widetilde S_{s,2}(Q) = \sum_{\substack{ R = R^T \\ R_{ii}\in \mathbb{Q}/\mathbb{Z},\, R_{ij}\in \mathbb{Q}}} \nu(R)^{-2s}e^{2\pi i \Tr(QR)},
\end{equation}
where the sum over the off-diagonal component of $R$ is over all rational numbers instead of just rational numbers mod one. The matrix $Q$ is given in terms of the charges by
\begin{equation}
	Q = \mat{\half(Q_1^2-\bar Q_1^2)}{\half(Q_1\cdot Q_2 - \bar Q_1 \cdot \bar Q_2)}{\half(Q_1\cdot Q_2 - \bar Q_1 \cdot \bar Q_2)}{\half(Q_2^2-\bar Q_2^2)}.
\end{equation}
The delta functions in (\ref{eq:fixedChargeTwoPoint}) constrain the charges to live on the surface of $(D-1)$-spheres with radii $\sqrt{\Delta_i\pm j_i}$. Integrating (\ref{eq:fixedChargeTwoPoint}) over these spheres gives the following simpler expression for the rank-two part of the two-point function of the density of states
\begin{equation}
\begin{aligned}\label{eq:rankTwoDensityDensity2}
	\langle\rho^p_{j_1}(\Delta_1)\rho^p_{j_2}(\Delta_2)\rangle_2 =& {4\pi^{4s}\over\Gamma\left(s\right)^4}\left((\Delta_1^2-j_1)^2(\Delta_2^2-j_2^2)\right)^{s-1}\\
	&\times \sum_{r_1,r_2,r\in\mathbb{Q}/\mathbb{Z}}\nu(R)^{-2s}e^{2\pi i(j_1 r_1+j_2r_2)}\sum_{n\in\mathbb{Z}}\prod_{\epsilon=\pm }\ofone\left(s;-\pi^2 (r+n)^2(\Delta_1-\epsilon j_1)(\Delta_2-\epsilon j_2)\right).
\end{aligned}
\end{equation}
That (\ref{eq:rankTwoDensityDensity}) and (\ref{eq:rankTwoDensityDensity2}) are equal can easily be checked numerically and appears to be a nontrivial integral identity. For $j_1,j_2$ both non-vanishing, the $n=r=0$ term above gives the entire disconnected contribution to the density-density correlator. One can also derive (\ref{eq:rankTwoDensityDensity2}) by summing over a particular genus-two modular crossing kernel, which we explain in appendix \ref{app:genusTwoCrossingKernels}.

Combining the rank-one and -two results, we can write the entire connected two-point function of the density of states for non-vanishing spins $j_1,j_2$ as
\begin{equation}
\begin{aligned}\label{eq:connectedDensityDensity}
	&\langle \rho^p_{j_1}(\Delta_1)\rho^p_{j_2}(\Delta_2)\rangle_{\rm conn}\\
	=& \, \sum_{j\ne 0}\sum_{\substack{(m,n)=1 \\ m^2 j = j_1,\, n^2 j = j_2}}\left\langle \rho^p_j\left({\Delta_1\over 2m^2}+{\Delta_2\over 2n^2}\right)\right\rangle\delta(n^2\Delta_1-m^2\Delta_2)\\
	& \, + {|j_1j_2|^{2s-1}\zeta(2s)^2\over\sigma_{2s-1}(j_1)\sigma_{2s-1}(j_2)}\langle \rho^p_{j_1}(\Delta_1)\rangle\langle \rho^p_{j_2}(\Delta_2)\rangle\sideset{}{'}\sum_{\substack{r_1,r_2\in\mathbb{Q}/\mathbb{Z}\\ r\in \mathbb{Q}}}\nu(R)^{-2s}e^{2\pi i (j_1r_1+j_2r_2)}\prod_{\epsilon=\pm}\ofone \left(s;-\pi^2 r^2(\Delta_1-\epsilon j_1)(\Delta_2-\epsilon j_2)\right),
\end{aligned}
\end{equation}
where the prime on the summation indicates that the $r=0$ term, corresponding to the disconnected correlator $\langle\rho^p_{j}(\Delta_1)\rangle\langle\rho^p_{j}(\Delta_2)\rangle$, is omitted. In the case that the spins are equal, $j_1 = j_2 = j$, the first line simplifies to $\left\langle \rho^p_j\left({\Delta_1+\Delta_2\over 2}\right)\right\rangle\delta(\Delta_1-\Delta_2)$. Similarly, in the scalar sector we have
\begin{equation}
\begin{aligned}\label{eq:connectedScalarDensityDensity}
	\langle\rho^p_0(\Delta_1)\rho^p_0(\Delta_2)\rangle_{\rm conn} =& \, \sideset{}{'}\sum_{(m,n)=1}\Big\langle \tilde\rho^p_0\left({\Delta_1\over 2m^2}+{\Delta_2\over 2n^2}\right)\Big\rangle\delta(n^2\Delta_1-m^2\Delta_2)\\
	& \, + {\zeta(2s)^2\over \zeta(2s-1)^2}\langle\tilde\rho^p_0(\Delta_1)\rangle\langle\tilde\rho^p_0(\Delta_2)\rangle\sideset{}{'}\sum_{\substack{r_1,r_2\in\mathbb{Q}/\mathbb{Z}\\ r\in\mathbb{Q}}}\nu(R)^{-2s}\ofone(s;-\pi^2 r^2\Delta_1\Delta_2)^2,
\end{aligned}
\end{equation}
where the prime on the summation over $m,n$ indicates that we are omitting $(m,n) = (1,0)$ and $(0,1)$ which contribute to the disconnected two-point function, and we have defined the modified scalar density $\langle \tilde\rho^p_0(\Delta)\rangle$ to be the scalar density of states (\ref{eq:narainDensityOfStates}) with the delta function representing the identity operator omitted
\begin{equation}
	\langle\tilde\rho^p_0(\Delta)\rangle = {2\pi^{2s}\zeta(2s-1)\over \Gamma(s)^2\zeta(2s)}\Delta^{2s-2}.
\end{equation}

It is worth commenting briefly on the interpretation of this two point function; we will consider the scalar two point function (\ref{eq:connectedScalarDensityDensity}) for simplicity.  We first note that the terms in the first line of (\ref{eq:connectedScalarDensityDensity}) have a simple intuitive explanation: the existence of a state with dimension $\Delta_1$ is perfectly correlated with the existence of a state with dimension equal to a rational number times $\Delta_1$.  Since we are considering ensembles of free bosons, this is no surprise: dimensions of primary operators with different windings/momenta will be related to one another by rational multiples.  The second line of equation  (\ref{eq:connectedScalarDensityDensity}) is more interesting, as that is the term which describes non-trivial correlations in the structure of spectrum. In a random matrix theory, for example, this term would be given by the sine kernel which describes (among other things) level repulsion \cite{Guhr:1997ve,Livan_2018}.  In the present case, however, we have found something rather different -- the statistics are not those of a random matrix, nor are they simply the Poisson statistics expected of a conventional integrable system.  
This unusual behaviour will be reflected in our discussion of the spectral form factor below. 

\subsection{The semiclassical limit}
In this section we will study the behaviour of the two-point function $\langle Z^p(iy_1)Z^p(iy_2)\rangle$ in the limit where the number of free bosons is taken to be large. One expects that in the $s\to\infty$ limit, the connected contributions to the two-point function are non-perturbatively suppressed. Here we will investigate the extent to which this is the case.

At fixed inverse temperature $y$, the large-$s$ limit of the one-point function is approximated by
\begin{equation}
	\langle Z^p(iy)\rangle \approx 1 + y^{-2s}.
\end{equation}
For simplicity in what follows we will set the inverse temperatures to be equal, $y_1 = y_2 = y$. In the low-temperature regime with $y>1$, the connected part of the two-point function is dominated by the $(m,n) = (1,1)$ rank-one term and the disconnected part is dominated by the rank-zero contribution so we have
\begin{equation}
	{\langle Z^p(iy)Z^p(iy)\rangle_{\rm connected}\over \langle Z^p(iy)Z^p(iy)\rangle_{\rm disconnected}} \approx (2y)^{-2s}.
\end{equation}
On the other hand, at high temperatures $y<1$ the disconnected part is approximated by
\begin{equation}
	\langle Z^p(iy)Z^p(iy)\rangle_{\rm disconnected} \approx y^{-4s},
\end{equation}
and the dominant connected contribution comes from the rank-two part of the two-point function. In particular, the largest connected contributions are the terms in the sum in (\ref{eq:eisensteinRankTwo}) for which $R=0$ and $\det(N) = 0$, so that
\begin{equation}
	{\langle Z^p(iy)Z^p(iy)\rangle_{\rm connected}} \approx 4(4y^2+y^4)^{-s}.
\end{equation}
So again the relative contribution to the two-point function of the connected terms is exponentially suppressed at large $s$ provided $y$ is sufficiently small. 

So far, we have only shown that the connected two-point function is exponentially suppressed when the inverse temperature are equal, in the low- and high-temperature limits.  It is possible that at generic values of $y_1,y_2$, there is no such exponential suppression.\footnote{We thank A. Dymarsky for discussions on this point.}

One may also ask whether connected configurations make an important contribution to the two-point function of the density of states in the large $s$ limit. It is evident from (\ref{eq:rankTwoDensityDensity2}) that the terms with $r\ne 0$ are exponentially suppressed relative to the disconnected correlator due to the factor of $\nu(R)^{-2s}$. However it is not a priori obvious that the terms with $r = 0$ and $n\ne 0$ are suppressed. Indeed, the connected correlator contains the following contribution
\begin{equation}
\begin{aligned}
	\langle \rho^p_{j_1}(\Delta_1)\rho^p_{j_2}(\Delta_2)\rangle_{\rm conn} &\supset 2\langle \rho^p_{j_1}(\Delta_1)\rangle\langle\rho^p_{j_2}(\Delta_2)\rangle \sum_{n=1}^\infty \prod_{\epsilon = \pm}\ofone(s;-\pi^2 n^2(\Delta_1-\epsilon j_1)(\Delta_2-\epsilon j_2))\\
	&\equiv \reallywidetilde{\langle \rho^p_{j_1}(\Delta_1)\rho^p_{j_2}(\Delta_2)\rangle}_{\rm conn}
\end{aligned}
\end{equation}
These contributions to the density-density correlator are actually \emph{not} exponentially suppressed at large $s$ compared to the disconnected part until the twists are sufficiently large, in particular until
\begin{equation}\label{eq:rhorhoTwistBound}
	\Delta_i - |j_i| \gtrsim {s\over \pi e}.
\end{equation}
As discussed at the end of section \ref{sec:partitionFunction}, this is precisely when the averaged density of states $\langle \rho^p_{j}(\Delta)\rangle$ ``turns on.'' To see this, we rewrite
\begin{equation}\label{eq:rhorhoTilde}
	{\reallywidetilde{\langle \rho^p_{j_1}(\Delta_1)\rho^p_{j_2}(\Delta_2)\rangle}_{\rm conn}\over \langle \rho^p_{j_1}(\Delta_1)\rangle\langle \rho^p_{j_2}(\Delta_2)\rangle} = {2\Gamma(s)^2\over \left(4\pi^2\sqrt{(\Delta_1^2-j_1^2)(\Delta_2^2-j_2^2)}\right)^{s-1}}\sum_{n=1}^\infty\prod_{\epsilon=\pm}{J_{s-1}\left(2\pi n\sqrt{(\Delta_1-\epsilon j_1)(\Delta_2-\epsilon j_2)}\right)\over n^{2s-2}}.
\end{equation}
In the large $s$ limit with fixed argument $z$, the Bessel function $J_{s-1}(z)$ is factorially small, and one can show that the combination (\ref{eq:rhorhoTilde}) is not exponentially suppressed. However, once the twists surpass the bound (\ref{eq:rhorhoTwistBound}), the relative contribution of the disconnected part becomes exponentially suppressed. A similar exponential suppression of the connected two-point function of the density of states in the large central charge limit was observed in the ensembles of chiral free bosons and of code CFTs \cite{Dymarsky:2020qom}.

\subsection{Comparison to the $T^2\times I$ wormhole amplitude}\label{subsec:torusWormhole}
By studying the properties of the degree-two Eisenstein series, have seen that the Siegel-Weil formula allows us to compute the second moment of the torus partition function in the Narain ensemble quite explicitly:
\begin{equation}\label{eq:fullTwoPointFunction}
	\langle Z^p(\tau_1)Z^p(\tau_2)\rangle = 1 + \sum_{(m,n)=1}\sum_{\substack{(c,d)=1 \\ c> 0}}|c(m^2\tau_1+n^2\tau_2)+d|^{-2s} + \sum_{\substack{P = P^T \\ \rm rational}}\nu(P)^{-2s}|\det\left(\Omega+P\right)|^{-2s},
\end{equation}
where $\Omega = \diag(\tau_1,\tau_2)$. However, as explained in the beginning of section \ref{sec:twoPoint}, the gravitational interpretation of the terms in the genus-two modular sum is somewhat indirect. In this section we will describe how our result compares to Euclidean wormhole amplitudes that have previously appeared in the literature, which will allow us to give some of the terms in the sum a conventional gravitational interpretation in terms of a simple class of Euclidean wormholes.

Recently, the path integral of three-dimensional Einstein gravity with negative cosmological constant on a simple class of Euclidean wormhole geometries topologically equivalent to a torus times an interval was considered \cite{Cotler:2020ugk}. Such configurations are not saddle points of the Einstein action, and so these wormhole amplitudes were computed somewhat indirectly using the first-order formalism of general relativity, and have since been discussed as examples of a sort of gravitational constrained instanton \cite{Cotler:2020lxj}. The result takes the form of a particular seed preamplitude summed over relative modular transformations of one of the boundary complex structures with respect to the other. The resulting amplitude exhibits features characteristic of double-scaled random matrix theory in the low-temperature limit, and leads to a linear ramp at late times in the spectral form factor, which is obtained from the wormhole amplitude by analytic continuation. In follow-up work \cite{Cotler:2020hgz}, the preamplitude was rederived from a sort of modular bootstrap,\footnote{Recent work \cite{Das:2021ipx} has applied these techniques to study torus wormhole amplitudes in higher-spin AdS$_3$ gravity, showing that in sectors of fixed spin at low temperatures, the spectral form factor exhibits a non-linear ramp at late times.} and similar reasoning was also used to compute the path integral $U(1)$ gravity on such $T^2 \times I$ wormholes, with the following result\footnote{In writing this formula we have multiplied by an overall factor of $|\eta(\tau_1)\eta(\tau_2)|^{4s}$ to facilitate a comparison with the second moment of the partition function of $U(1)^D\times U(1)^D$ primary operators.}
\begin{equation}
\begin{aligned}\label{eq:torusInterval}
	\mathcal{Z}^{\rm Narain}_{T^2\times I}(\tau_1,\tau_2) &= (\im(\tau_1)\im(\tau_2))^{-s} \sum_{\gamma\in PSL(2,\mathbb{Z})} \left({\im(\tau_1)\im(\gamma\tau_2)\over|\tau_1+\gamma\tau_2|^2}\right)^s \\
	& \equiv \left(\im(\tau_1)\im(\tau_2)\right)^{-s} \sum_{\gamma\in PSL(2,\mathbb{Z})} \left(\mathcal{Z}_0(\tau_1,\gamma\tau_2)\right)^s.
\end{aligned}
\end{equation}
The seed amplitude $\mathcal{Z}_0^s$ is interpreted in \cite{Cotler:2020hgz} as the gravitational path integral of $U(1)$ gravity on the three-manifold $T^2\times I$ (with no Dehn twist between the two boundaries) with boundary modular parameters $\tau_1,\tau_2$, and is invariant under the following simultaneous modular transformation
\begin{equation}\label{eq:peculariarModularInvariance}
	\mathcal{Z}_0\left(\gamma\tau_1, (M\gamma M)\tau_2\right) = \mathcal{Z}_0(\tau_1,\tau_2),\quad M = \mat{-1}{0}{0}{1},
\end{equation}
where $M$ takes $\tau \rightarrow - \tau$. The sum over $PSL(2,\mathbb{Z})$ images in (\ref{eq:torusInterval}) is interpreted as a sum over Dehn twists of one boundary with respect to the other, which restores invariance with respect to independent modular transformations of the two boundaries. In what follows it will be instructive to rewrite the amplitude (\ref{eq:torusInterval}) as follows
\begin{equation}\label{eq:torusInterval2}
	\mathcal{Z}^{\rm Narain}_{T^2\times I}(\tau_1,\tau_2) = \sum_{\gamma\in PSL(2,\mathbb{Z})}|c'\tau_1\tau_2+d'\tau_1 + a'\tau_2 + b'|^{-2s},\quad \gamma=\mat{a'}{b'}{c'}{d'}\in PSL(2,\mathbb{Z}).
\end{equation}

Some of the terms in the genus-two modular sum appearing in the full two-point function (\ref{eq:fullTwoPointFunction}) are realized in the $T^2\times I$ amplitude (\ref{eq:torusInterval}), and so can be associated with this simple class of Euclidean wormhole configurations. In particular, the rank-one terms in (\ref{eq:fullTwoPointFunction}) with $(m,n) = (1,1)$ and $c=1$ with $d\in\mathbb{Z}$ map onto the $PSL(2,\mathbb{Z})$ images in (\ref{eq:torusInterval2}) that shift $\tau_2$ by an integer, namely those with $c'=0, b'\in\mathbb{Z}$. In the case of Einstein gravity originally studied in \cite{Cotler:2020ugk}, these were precisely the terms that were dominant at low temperatures and led to the linear ramp in the spectral form factor. In section \ref{sec:SFF} we will see that they contribute to the late-time plateau in $U(1)$ gravity.\footnote{Already in \cite{Maloney:2020nni} and \cite{Cotler:2020hgz} it was noticed that these terms contribute a constant to the late-time behaviour of the spectral form factor.} On the other hand, the terms in (\ref{eq:torusInterval2}) with $c'\ne 0$ come from the rank-two terms in (\ref{eq:fullTwoPointFunction}). Writing $P = \mat{p_1}{p}{p}{p_2}$, we have
\begin{equation}
	\langle Z^p(\tau_1)Z^p(\tau_2)\rangle_2 = \sum_{\substack{P = P^T \\ \rm rational}}\nu(P)^{-2s}|\tau_1\tau_2 + p_2\tau_1 + p_1\tau_2 + p_1p_2-p^2|^{-2s}
\end{equation}
and so the $T^2 \times I$ wormholes correspond to the special cases
\begin{equation}
	p_1 = {a'\over c'},\quad p_2 = {d'\over c'}, \quad p = {1\over c'}.
\end{equation}
In other words, they correspond to $Sp(4,\mathbb{Z})$ elements with lower entries
\begin{equation}
\begin{aligned}
  C = C_\gamma &= \begin{pmatrix} 0 & c' \\ -1 & a' \end{pmatrix}\\
  D = D_\gamma &= \begin{pmatrix} 1 & d' \\ 0 & b' \end{pmatrix}
\end{aligned}
\end{equation}
Indeed, from (\ref{eq:contractibleCycles}) we see that they correspond to bulk topologies $M_{C_\gamma,D_\gamma}$ with the following contractible cycles (recall from the discussion at the beginning of section \ref{sec:twoPoint} that $A_i$ and $B_i$ are the $A$- and $B$-cycles of the boundary tori)
\begin{equation}
\begin{aligned}
  M_{C_\gamma,D_\gamma}:~~\widetilde A_1 &= A_1 + c'B_2 + d'A_2 \\
  \widetilde A_2 &= -B_1 + a'B_2 + b' A_2.
\end{aligned}
\end{equation}

The other terms in the genus-two modular sum in (\ref{eq:fullTwoPointFunction}) cannot straightforwardly be associated with $T^2\times I$ wormholes.

We now briefly pause to compare our result to the path integral of pure 3D gravity on torus wormholes as computed in \cite{Cotler:2020ugk}. The pure gravity wormhole amplitude was found to be\footnote{Below we have again multiplied the result by an overall factor of $|\eta(\tau_1)\eta(\tau_2)|^{2}$ so that the result captures information about the second moment of the partition function of Virasoro primary operators.}
\begin{equation}\label{eq:pureGravityTorusInterval}
	\mathcal{Z}_{T^2\times I}^{\text{pure gravity}}(\tau_1,\tau_2) = \left(\im(\tau_1)\im(\tau_2)\right)^{-\half}\sum_{\gamma\in PSL(2,\mathbb{Z})}\mathcal{Z}_0(\tau_1,\gamma\tau_2).
\end{equation}
That the low-temperature spectral form factor inherited from (\ref{eq:pureGravityTorusInterval}) exhibits a linear ramp rather than a plateau at late Lorentzian times is essentially due to the relative difference between the powers of $\im(\tau_1)\im(\tau_2)$ and of the preamplitude $\mathcal{Z}_0$. On the other hand, these powers cancel in the torus wormhole amplitude in the Narain ensemble (\ref{eq:torusInterval}), leading to a plateau at late times. Indeed as we will see in section \ref{sec:SFF}, the spectral form factor obtained by analytic continuation of the full two-point function of primaries (\ref{eq:fullTwoPointFunction}) exhibits a plateau at late times.

\section{Higher-point functions}\label{sec:higherPoint}
In this section we will outline the computation of higher moments of the torus partition function. 
The averaged $n$-point function is  computed by a degree $n$ Eisenstein series 
\begin{equation}
\begin{aligned}
	\langle Z^p(\tau_1)\cdots Z^p(\tau_n)\rangle &= (\det\im\Omega)^{-s}E_s^{(n)}(\Omega)
	&= \sum_{(C,D)=1}|\det(C\Omega+D)|^{-2s},
\end{aligned}
\end{equation}
where the modular parameters $\tau_i$ are packaged into a diagonal period matrix
\begin{equation}
	\Omega = \diag(\tau_1,\ldots,\tau_n)
\end{equation}
and the sum over matrices $(C,D)$ is over symmetric coprime pairs as described in appendix \ref{app:eisensteinFourier}. As in the case of the two-point function, we will write this as
\begin{equation}
	\langle Z^p(\tau_1)\cdots Z^p(\tau_n)\rangle = 1 + \langle Z^p(\tau_1)\cdots Z^p(\tau_n)\rangle_1 + \ldots + \langle Z^p(\tau_1)\cdots Z^p(\tau_n)\rangle_n
\end{equation}
where $\langle Z^p(\tau_1) \cdots Z^p(\tau_n)\rangle_r$ denotes the sum of contributions from terms where $C$ has rank $r$.

Following the discussion in appendix \ref{app:eisensteinFourier}, a complete set of representatives of coprime symmetric pairs $\{C,D\}$ where $C$ has rank $r$ is given by a pair of coprime $r\times r$ matrices $\{C^{(r)},D^{(r)}\}$ with $\det C^{(r)}\ne 0$, along with a unimodular $n\times n$ matrix $U$
\begin{equation}
\begin{aligned}
	C &= \mat{C^{(r)}}{0}{0}{0}U^T\\
	D &= \mat{D^{(r)}}{0}{0}{\id}U^{-1}.
\end{aligned}
\end{equation}
The set of such matrices $U$ is given by the set of primitive $n\times r$ matrices $Q^{(n,r)}$ that can be completed to a unimodular matrix $U$, subject to the equivalence relation $Q^{(n,r)}\sim Q^{(n,r)}U^{(r)}$ where $U^{(r)}$ is a unimodular $r\times r$ matrix.

The upshot of this decomposition is that the contribution to the correlation function where $\rank C = r$, $\langle Z^p(\tau_1)\cdots Z^p(\tau_n)\rangle_r$, can be written in terms of degree $r$ Eisenstein series. We already saw an example of this with the rank-one contributions to the two-point function studied in section \ref{subsec:rankOne}, which involved a sum over genus one Eisenstein series. 

A particularly tractable example is the rank-one contributions to the $n$-point function. In this case the complete set of representatives $\{Q^{(n,1)}\}$ is given by
\begin{equation}
	Q^{(n,1)} = (m_1\cdots  m_n)^T~ \text{with $\gcd(m_1,\ldots,m_n)=1$}.
\end{equation}
Defining
\begin{equation}
	\tau_{m_1,\cdots,m_n} = \sum_{i=1}^n m_i^2\tau_i,
\end{equation}
we can write the rank-one contribution to the $n$-point function in terms of a sum of one-point functions:
\begin{equation}
\begin{aligned}\label{eq:nPointRank1}
	\langle Z^p(\tau_1)\cdots Z^p(\tau_n)\rangle_1 &= \sum_{\gcd(m_1,\ldots,m_n)=1}\sum_{\substack{(c,d)=1\\c > 0}}|c\tau_{m_1,\cdots m_n}+d|^{-2s}\\
	&= \sum_{\gcd(m_1,\ldots m_n)=1}\left(\langle Z^p(\tau_{m_1,\cdots m_n})\rangle - 1\right).
\end{aligned}
\end{equation}

Similarly, following the discussion in appendix \ref{app:eisensteinFourier}, the rank-$r$ contributions to the $n$-point function with $1<r<n$ can be written in terms of a sum over the maximal-rank contributions to the $r$-point function with modified values of the worldsheet moduli. The maximal rank contributions to the $n$-point function are simply given by the following sum over symmetric rational $n\times n $ matrices
\begin{equation}
	\langle Z^p(\tau_1)\cdots Z^p(\tau_2)\rangle_n = \sum_{\substack{P = P^T \\ \text{rational}}} \nu(P)^{-2s}|\det(\Omega+P)|^{-2s}.
\end{equation}
We leave a more detailed study of these higher multi-boundary observables in $U(1)$ gravity to future work.

\section{The spectral form factor}\label{sec:SFF}

The spectral form factor has emerged as an important diagnostic of chaos and probe of spectral statistics for complex systems. For a quantum system at finite temperature $\beta^{-1}$ with Hamiltonian $H$, the spectral form factor is 
\begin{equation}\label{eq:SFF}
	g(\beta,T) = |\Tr(e^{-(\beta+iT)H})|^2 = \sum_{i,j}d_{i}d_j e^{-\beta(E_i+E_j)+iT(E_i-E_j)},
\end{equation}
where the sum over $i$ is over states in the Hilbert space of the quantum system and $d_i$ labels the degeneracy of the state with energy $E_i$. This quantity is interesting in part because it is a simple surrogate for the thermal two-point function of local operators \cite{Cotler:2016fpe}, which in a holographic system captures the two-point correlation function of bulk quantum fields outside a black hole horizon separated by some Lorentzian time $T$. In Maldacena's formulation of the information problem for eternal black holes in AdS \cite{Maldacena:2001kr}, this quantity plays a central role: it decays exponentially at late times in semiclassical gravity due to quasinormal mode relaxation, but this is forbidden in a unitary quantum system with a discrete Hilbert space. Indeed, the late-time average of (\ref{eq:SFF}) is bounded from below
\begin{equation}\label{guillam}
	\lim_{T\to\infty} {1\over T}\int_0^T dt\, g(\beta,t) \ge \Tr e^{-2\beta H}.
\end{equation}
In a chaotic system in which there are typically no degeneracies, this bound is saturated. In this way the spectral form factor captures the tension between the smooth geometry of semiclassical gravity and the discrete Hilbert space of black hole microstates. 

Furthermore, it has been conjectured \cite{Cotler:2016fpe} that in theories with quantum mechanical black holes, the spectral form factor follows the universal behaviour prescribed by random matrix theory, including the late-time plateau described above that is a signature of discreteness of the spectrum. This behaviour is additionally characterized by an early-time decay below the plateau value, followed by a period of linear growth (sometimes called the ``ramp'') approaching the plateau from below. In random matrix theory, the linear ramp is a consequence of eigenvalue repulsion.

In this section we will study the spectral form factor in the Narain ensemble of free conformal field theories, which is obtained by analytic continuation of the two-point correlation function of the torus partition function of primaries (\ref{eq:twoPointFunction}) to $y_1 = \beta+iT$, $y_2 = \beta-iT$ with $x_1=x_2=0$:
\begin{equation}
	g(\beta,T) = \langle Z^p(\beta+iT)Z^p(\beta-iT)\rangle.
\end{equation} 
We emphasize that in expressions like this the argument of the partition function is meant to denote the imaginary part of the modular parameter analytically continued to real time, but that the real part of the modular parameter is set to zero.\footnote{Although of course we could consider the spectral form factor in sectors of definite spin, which would involve the Fourier transform with respect to the real part of the modular parameter.}
We will discover that the ``wormhole sum" for the Narain ensemble precisely reproduces the Plateau behaviour anticipated in equation (\ref{guillam}).

Using our results from section \ref{subsec:rankOne}, the rank-one contributions to the spectral form factor can be written as
\begin{equation}
\begin{aligned}\label{eq:SFFRk1}
	\langle Z^p(\beta+iT)Z^p(\beta-iT)\rangle_1 =& \, \langle Z^p(2\beta) \rangle - 1\\
	& \, + 2\sideset{}{'}\sum_{(m,n)=1}\bigg[b_s(0)(\beta^2_{m,n}+T^2_{m,n})^{\half-s}\cos( (2s-1)\phi_{m,n}))\\
	& \, +\sum_{j=1}^\infty b_s(j)\bigg((\beta_{m,n}+iT_{m,n})^{\half-s}K_{s-\half}(2\pi j(\beta_{m,n}+i T_{m,n}))\\
	& \, +(\beta_{m,n}-iT_{m,n})^{\half-s}K_{s-\half}(2\pi j(\beta_{m,n}-i T_{m,n}))\bigg)\bigg],
\end{aligned}
\end{equation}
where 
\begin{equation}
\begin{aligned}
	\beta_{m,n} &\equiv (m^2+n^2)\beta\\
	T_{m,n} &\equiv (m^2-n^2)T\\
	\phi_{m,n} & \equiv \arctan\left({T_{m,n}\over \beta_{m,n}}\right)
\end{aligned}
\end{equation}
and the prime on the summation over coprime integers $(m,n)$ indicates that we are only including pairs of coprime numbers for which $m>n$. In particular $(m,n) = (1,1)$ is excluded. The first line in (\ref{eq:SFFRk1}) is the late-time plateau, while the term with $m=1,n=0$ is the rank-one part of the disconnected correlator, namely $\langle Z^p(\beta+iT)\rangle +\langle Z^p(\beta-iT)\rangle - 2$. Meanwhile, the rank-two contributions to the spectral form factor are
\begin{equation}
\begin{aligned}\label{eq:SFFRk2}
	&\langle Z^p(\beta+i T)Z^p(\beta-iT)\rangle_2\\ 
	=& \sum_{n_1,n_2,n\in\mathbb{Z}}\sum_{r_1,r_2,r\in\mathbb{Q}/\mathbb{Z}}{\nu(R)^{-2s}\over\left(\widetilde n^4 + 2\widetilde n^2(\beta^2+T^2-\widetilde n_1\widetilde n_2)+(\widetilde n_1^2+(\beta+iT)^2)(\widetilde n_2^2+(\beta-iT)^2)\right)^s},
\end{aligned}
\end{equation}
where $\widetilde n,\widetilde n_1,\widetilde n_2$ are as defined in (\ref{eq:definedTildedNs}) and $\nu(R)$ is the product of the denominators of the elementary divisors of the rational symmetric matrix $R = \mat{r_1}{r}{r}{r_2}$. We remind the reader that the $n=r=0$ term contributes part of the disconnected part of the spectral form factor $(\langle Z^p(\beta+iT)\rangle-1)(\langle Z^p(\beta-iT)\rangle-1)$.

From the rank-one and rank-two contributions to the two-point function elucidated in section \ref{sec:twoPoint}, many features of the plot of the spectral form factor as a function of time $T$ are already apparent. 

\subsection*{Early time}
At early times, the behaviour of the spectral form factor depends sensitively on the temperature, and receives contributions from both the rank-one and rank-two parts of the two-point function as discussed in section \ref{sec:twoPoint}. For example, in the low-temperature ($\beta\to\infty$) limit, the rank-one contribution (\ref{eq:SFFRk1}) dominates:
\begin{equation}
\begin{aligned}
	g(\beta,0)-1
	&\sim \left(b_s(0)\sum_{(m,n)=1}(m^2+n^2)^{1-2s}\right)\beta^{1-2s}\\
	&= b_s(0)\left({1\over 2\zeta(4s-2)}Z(\mathbb{1},2s-1)+1\right)\beta^{1-2s}\\
	&= b_s(0)\left(\half E_{2s-1}(i)+1\right)\beta^{1-2s}\\
	&= b_s(0)\left[{\zeta(2s-1)\over 4^{2s-1}\zeta(4s-2)}\left(\zeta_{1/4}(2s-1)-\zeta_{3/4}(2s-1)\right)+1\right]\beta^{1-2s},
\end{aligned}
\end{equation}
where (for $\re s>1$)
$\zeta_a(s) = \sum_{k=0}^\infty (k+a)^{-s}$ is the Hurwitz zeta function and
\begin{equation}
	Z(Y,s) = \half\sideset{}{'}\sum_{a\in \mathbb{Z}^2} (a^T Y a)^{-s}
\end{equation}
is the Epstein zeta function, with the prime denoting the omission of the origin in the sum.
In the semiclassical limit of large $s$, the $(m,n)=(1,0),(0,1)$ terms give the dominant contributions in the low-temperature regime up to corrections that are non-perturbatively suppressed in $s$. In this case the dominant contribution is from the disconnected correlator $2(\langle Z^p(\beta)\rangle-1)\sim 2b_s(0)\beta^{1-2s}$.

On the other hand, from (\ref{eq:SFFRk2}) we see that in the high-temperature ($\beta\to 0$) limit the disconnected part of the rank-two contribution dominates
\begin{equation}
	g(\beta,0)-1 \sim (\langle Z^p(\beta)\rangle-1)^2 \sim \beta^{-4s}.
\end{equation}
This characterizes the high-temperature behaviour regardless of whether or not $s$ is also taken to be large. The subsequent decay is then well approximated by $(\beta^2+T^2)^{-2s}$.

\subsection*{Late time}
At late times, the amplitude of both the rank-one and rank-two contributions decay, with the exception of a single rank-one contribution --- namely, the term in (\ref{eq:twoPointFunctionRankOne}) with $(m,n) = (1,1)$, which is constant in time --- leading to a plateau at late time
\begin{equation}
	\lim_{T\to\infty} g(\beta,T) = \langle Z^p(2\beta)\rangle%\over \langle Z^p(\beta)Z^p(\beta)\rangle}.
	.
\end{equation}
From (\ref{eq:twoPointFunctionRankOne}) we can then explicitly identify the contributions responsible for the late-time plateau in the spectral form factor, which is normally regarded as a diagnostic for discreteness of the spectrum. In particular, it is the class of $Sp(4,\ZZ)$ transformations of the diagonal matrix $\Omega = \diag(\tau_1,\tau_2)$ with lower elements
\begin{equation}
	C = \mat{c}{c}{0}{0},~~~~~
	D = \mat{d}{0}{-1}{1},
\end{equation}
with $c,d$ non-negative coprime integers that lead to the plateau at late time. They correspond to bulk topologies $M_{C,D}$ with the following cycles contractible in the bulk (recall that $A_i$ and $B_i$ are the $A$- and $B$-cycles of the boundary tori)
\begin{equation}\label{eq:plateauCycles}
\begin{aligned}
  M_{C,D}:~~\widetilde A_1 &= d A_1 + c(B_1 + B_2)\\
  \widetilde A_2 &= -A_1 + A_2.
\end{aligned}
\end{equation} 
As noted earlier, the terms with $c=1$ and $d\in\mathbb{Z}$ can be associated with simple Euclidean wormholes topologically equivalent to a torus times an interval. The remaining terms needed to reconstruct $\langle Z^p(2\beta)\rangle$ (i.e. those with $c>1$) do not seem to admit such an interpretation; they must be associated with other three-manifolds with two boundary tori. It would be very interesting to explicitly identify the corresponding bulk topologies, and to study the path integral of three-dimensional pure Einstein gravity on these backgrounds.

\subsection*{Intermediate time}
\begin{figure}[h!]
	\centering
	{
	\subfloat{\includegraphics[width=.49\textwidth]{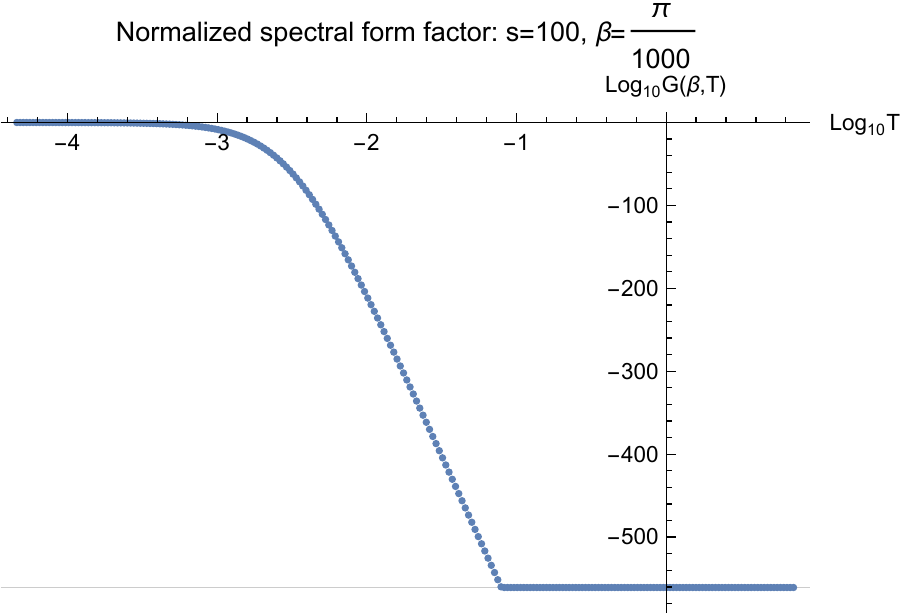}}~
	\subfloat{\includegraphics[width=.49\textwidth]{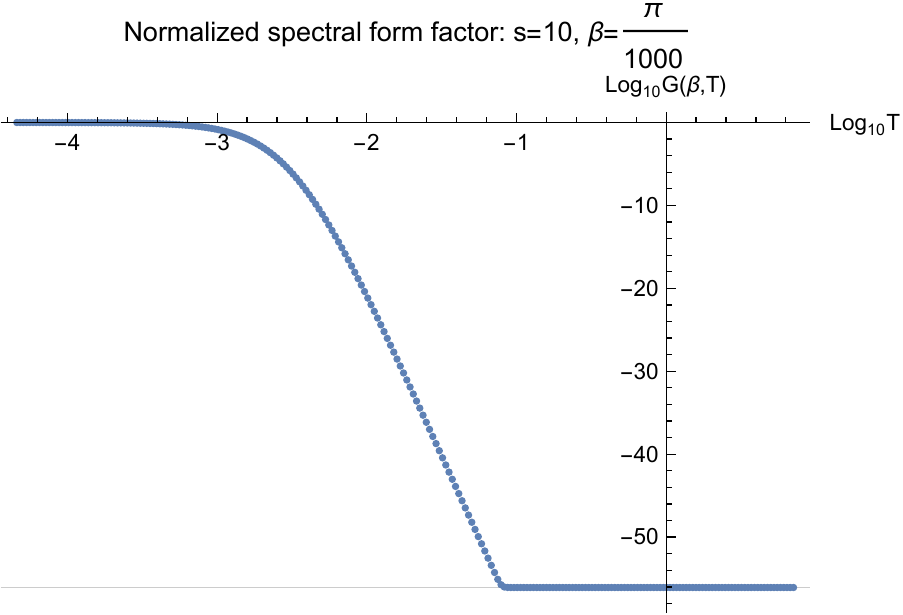}}\\
	\subfloat{\includegraphics[width=.49\textwidth]{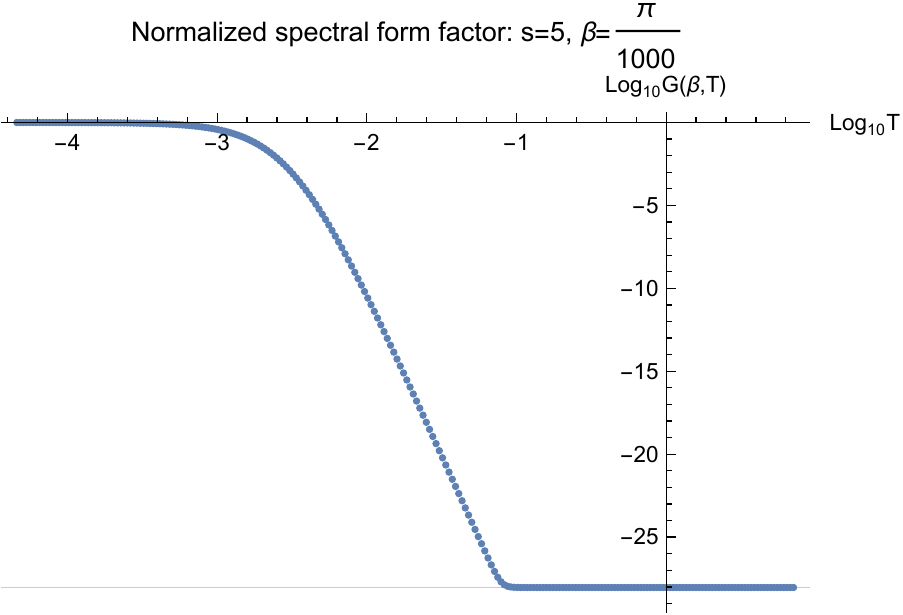}}
	}
	\caption{A plot of the normalized spectral form factor $ G(\beta,T) = g(\beta,T)/g(\beta,0)$ as a function of time in the averaged Narain CFT, for $s=100,10,5$ and with $\beta = {\pi\over 1000}$. The horizontal gridline corresponds to the plateau ${\langle Z^p(2\beta)\rangle \over \langle Z^p(\beta)^2\rangle}$. At smaller values of $\beta$, the rank-two contributions to the two-point function dominate at early times and decay monotonically, until a crossover time when they are surpassed by the plateau coming from the rank-one contributions.}\label{fig:smallBeta}
\end{figure}

The qualitative shape of the spectral form factor as a function of time depends sensitively on the value of the temperature. At high temperature, the rank-two contribution gives the dominant contribution at early times, of order $\langle Z^p(\beta)Z^p(\beta)\rangle_2 \approx (\langle Z^p(\beta)\rangle-1)^2 \sim \beta^{-4s}$. This decays with time until it passes below the plateau $\langle Z^p(2\beta)\rangle\sim(2\beta)^{-2s}$ from the rank-one contributions. Meanwhile, the rank-one contributions oscillate around the plateau with an amplitude that decays with time, as can be seen from (\ref{eq:SFFRk1}). Thus in this case it appears that the plateau is approached from above, up to the small oscillations around the plateau from the rank-one contributions. We compute the spectral form factor numerically by truncating the sums on integers and rational numbers in equations (\ref{eq:SFFRk1}) and (\ref{eq:SFFRk2}), and plot some examples of the high-temperature spectral form factor as a function of time in figure \ref{fig:smallBeta}.

\begin{figure}[h!]
	\centering
	{
	\subfloat{\includegraphics[width=.49\textwidth]{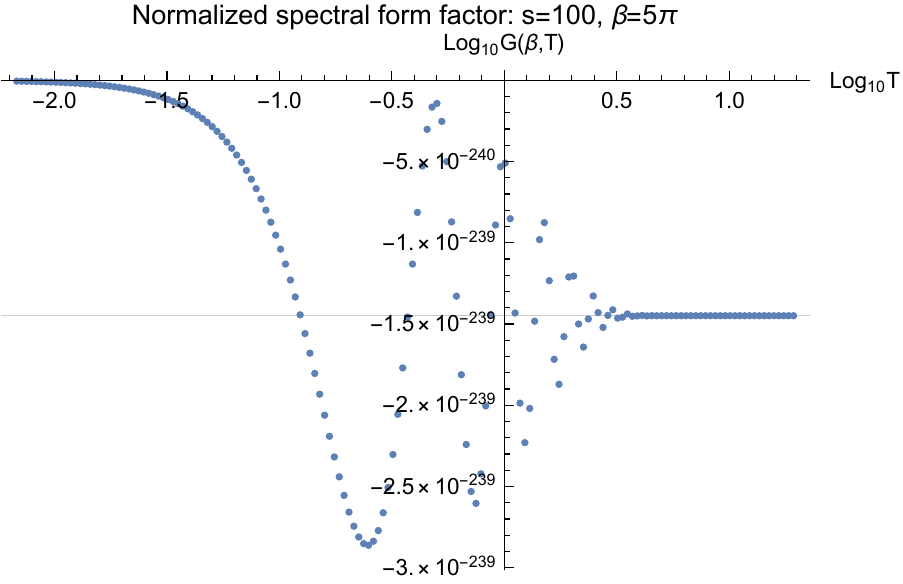}}~
	\subfloat{\includegraphics[width=.49\textwidth]{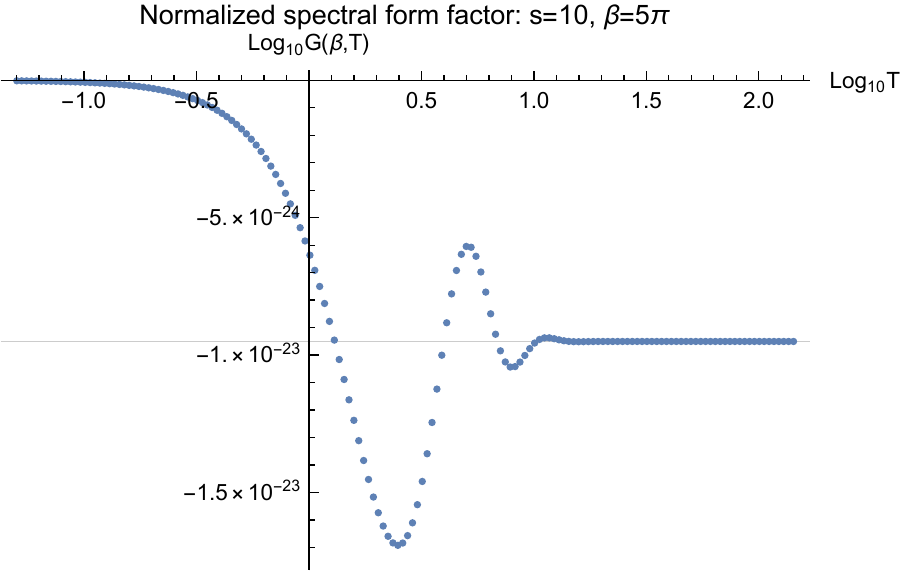}}\\
	\subfloat{\includegraphics[width=.49\textwidth]{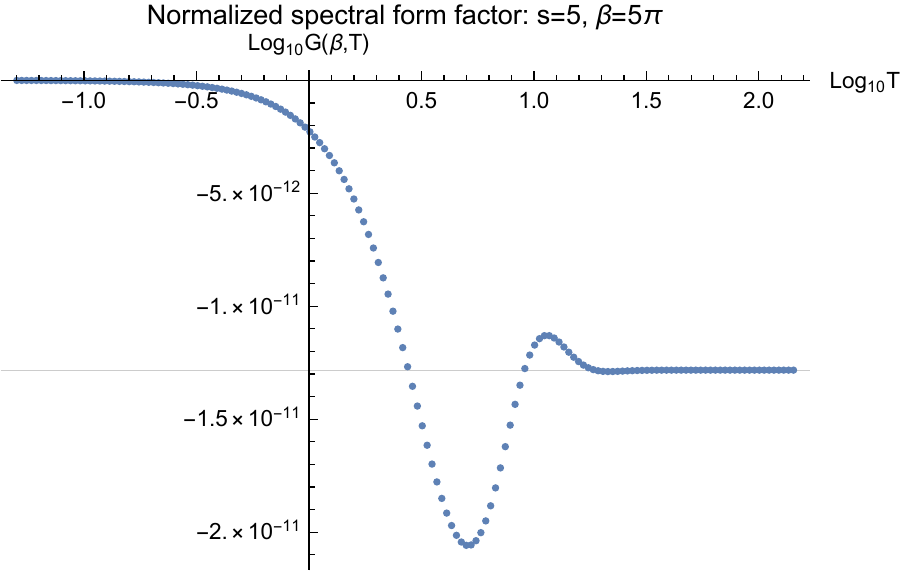}}
	}
	\caption{A plot of the normalized spectral form factor as a function of time in the averaged Narain CFT, for $s=100,10,5$ and with $\beta = {5\pi}$. At large values of $\beta$, the rank-two contributions to the two-point function are essentially never important, and the spectral form factor approaches the plateau in an oscillatory manner due to the rank-one contributions. The horizontal gridline corresponds to the plateau ${\langle Z^p(2\beta)\rangle \over \langle Z^p(\beta)^2\rangle}$.}\label{fig:largeBeta}
\end{figure}

At sufficiently low temperature, the rank-two terms essentially never give an important contribution to the spectral form factor. Aside from the rank-zero term (otherwise known as 1), the early time behaviour is dominated by the rank-one contribution $\langle Z^p(\beta)Z^p(\beta)\rangle_1\sim \beta^{1-2s}$. As one can see from (\ref{eq:SFFRk1}), as time increases the plateau is approached in an oscillatory manner with a decaying amplitude, so while the spectral form factor does ``dip'' below the plateau, it does not strictly approach the plateau from below. We again compute the spectral form factor numerically by truncation; some low-temperature examples are plotted in figure \ref{fig:largeBeta}.

\begin{figure}[h!]
	\centering
	{\subfloat{\includegraphics[width=.45\textwidth]{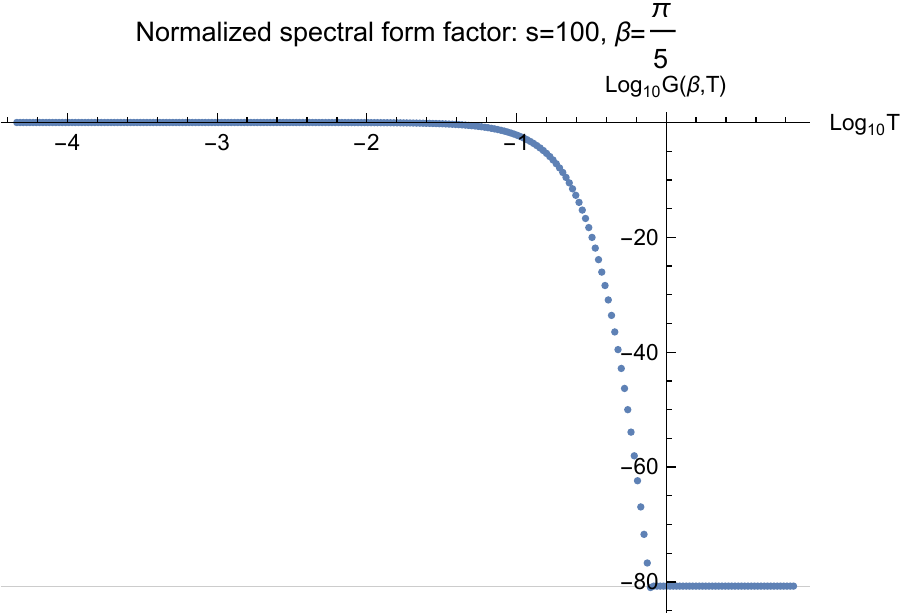}}~
	\subfloat{\includegraphics[width=.45\textwidth]{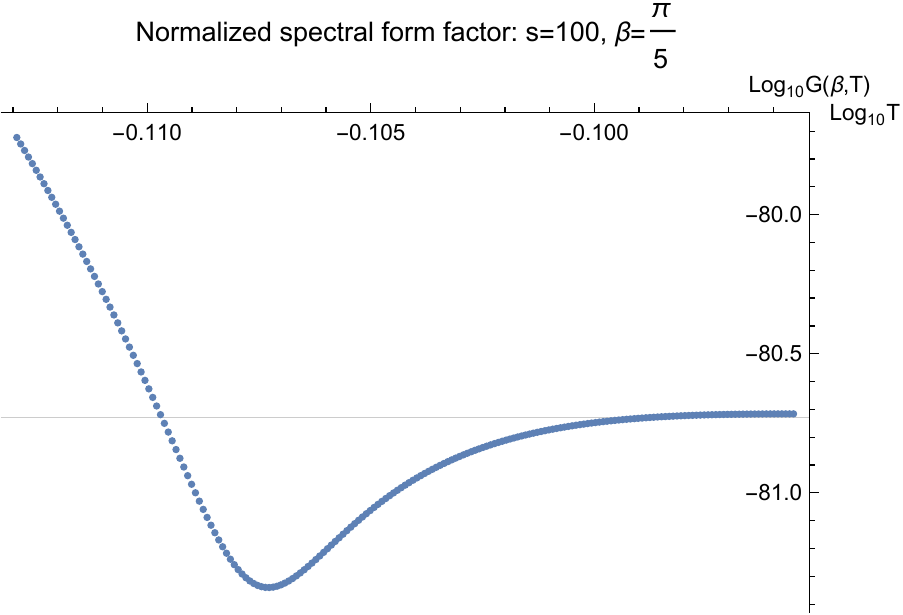}}}\\
	\subfloat{\includegraphics[width=.65\textwidth]{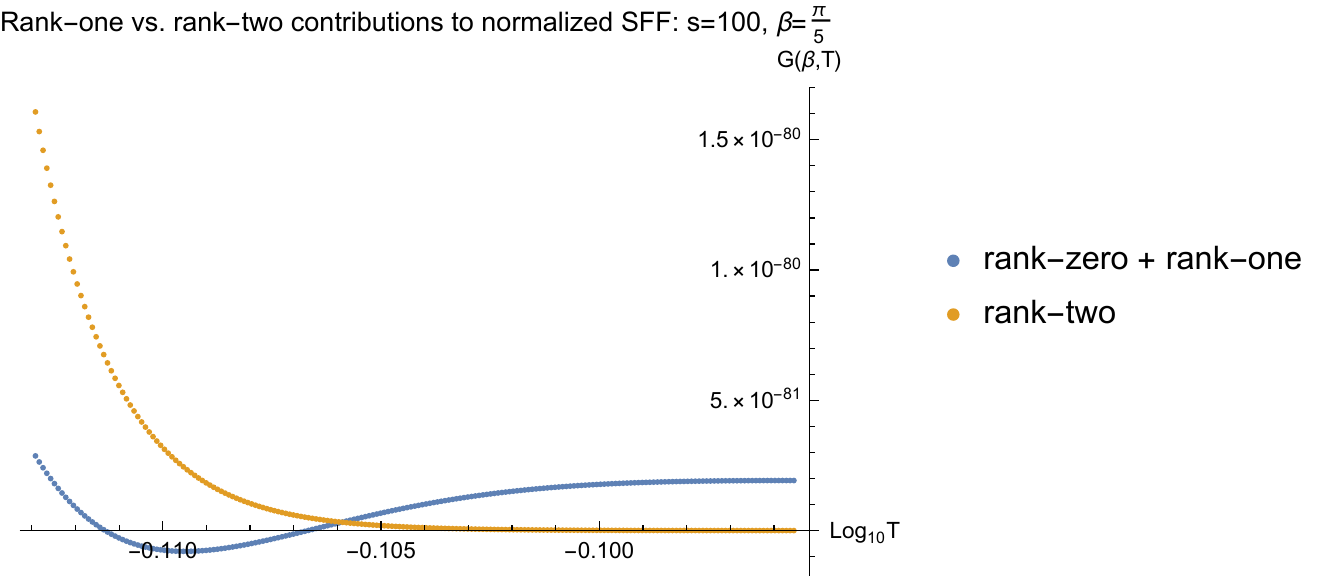}}
	\caption{A finely-tuned example of the normalized spectral form factor at a moderate value of the temperature $\beta = {\pi\over 5}$ with $s=100$. In the second plot on the right we zoom in to the region where the rank-one contributions overtake the rank-two contributions, demonstrating an apparent approach to the plateau from below. In the third plot at the bottom we show the separate rank-one and rank-two contributions to the spectral form factor.}\label{fig:mediumBeta}
\end{figure}

We have seen that the shape of the spectral form factor depends sensitively on the temperature, and in any case, we do not find a linear approach to the ramp from below that typically indicates long distance eigenvalue repulsion characteristic of random matrix theory. We do note, however, that there are values of the temperature where the spectral form factor approaches the ramp from below. In particular, the temperature can be tuned to a moderate value so that the rank-one contributions surpass the rank-two contributions at the bottom of an oscillation of the rank-one terms. We present an example of such a spectral form factor in figure \ref{fig:mediumBeta}, but emphasize that as in the high- and low-temperature scenarios the spectral form factor continues to oscillate around the plateau with an envelope that decays in time.

In the semiclassical regime, i.e. at large $s$, almost all of the interesting features of the spectral form factor at early and intermediate times are captured by the analytic continuation of the disconnected correlator $\langle Z^p(\beta+iT)\rangle \langle Z^p(\beta-iT)\rangle$, essentially until the onset of the late-time plateau. In this regime the dominant contribution to the connected correlator comes from the time-independent plateau itself.

\section{Discussion}
In this paper we have studied the spectral statistics of primary operators in the ensemble average of Narain's family of free boson CFTs by explicit computation of moments of the torus partition function in terms of higher-degree Eisenstein series. We obtained an exact expression for the two-point function of the density of primary states, and computed the spectral form factor of the bulk dual theory of ``$U(1)$ gravity'' by direct computation of an appropriate analytic continuation of a degree-two Eisenstein series. This two-boundary amplitude receives contributions from both disconnected and connected configurations in $U(1)$ gravity. While it was straightforward to identify the connected configurations that lead to the late-time plateau, there does not appear to be a linear ramp approaching the plateau from below, suggesting that the distribution of states in $U(1)$ gravity are not captured by random matrix theory. This is in contrast to a recent computation of the gravitational path integral in three-dimensional pure Einstein gravity on Euclidean wormhole backgrounds topologically equivalent to a torus times an interval \cite{Cotler:2020ugk,Cotler:2020hgz}, where it was argued that there is a regime of fixed angular momentum and low temperature where the spectral form factor exhibits a linear ramp at late times. In contrast, in our computation of the full connected two-point function of the density of primary states (see equations (\ref{eq:connectedDensityDensity}) and (\ref{eq:connectedScalarDensityDensity})) we did not find evidence of long-range repulsion of the scaling dimensions.

While we were able to identify the terms in the $Sp(4,\mathbb{Z})$ sum over wormholes that are responsible for reconstructing the late-time plateau of the spectral form factor of $U(1)$ gravity, we did not explicitly construct all of the corresponding topologies, partly due to the fact that $U(1)$ gravity is only sensitive to certain coarse information about the bulk. It would be very interesting to construct candidate topologies with the appropriate cycles contractible in the bulk (given by (\ref{eq:plateauCycles})) and study the path integral of pure AdS$_3$ Einstein gravity on these backgrounds.

We expect that the technology developed in this paper could find applications in the study of other higher-genus and multi-boundary observables in the bulk dual of the Narain ensemble average. One attainable target would be the free energy for $U(1)$ gravity with torus boundary computed via the replica trick. In particular, since the boundary theory is defined by an ensemble average, one can compute both the annealed and quenched free energy
\begin{equation}
\begin{aligned}
	F_{\rm annealed}(\beta) &= -\beta^{-1}\log\langle Z^p(\beta)\rangle\\
	F_{\rm quenched}(\beta) &= -\beta^{-1}\langle\log Z^p(\beta)\rangle,
\end{aligned}
\end{equation}
which are distinguished by whether one takes the ensemble average before or after computing the logarithm. In a recent study of the free energy in the $\widehat{\text{CGHS}}$ model and JT gravity \cite{Engelhardt:2020qpv}, it was emphasized that the annealed free energy is pathological and that connected topologies were an essential ingredient in the well-definedness of the quenched free energy at low temperature as computed from the gravitational path integral via the replica trick. Although in the Narain ensemble the annealed free energy is not obviously pathological (essentially due to the fact that the vacuum is a normalizable state), one could imagine similarly computing the quenched free energy in $U(1)$ gravity via the replica trick. However, the fact that for a fixed central charge $D$, the $n$-point function $\langle Z^p(\tau_1)\cdots Z^p(\tau_n)\rangle$ fails to converge for $n\ge D-1$ naively presents an obstacle to this computation via the replica trick. One might hope to get around this via the meromorphic continuation of the corresponding degree-$n$ Eisenstein series $E^{(n)}_s$, however these have poles at values of $s$ that prevent this from providing an straightforward resolution \cite{Kalinin_1978}. Nonetheless, in the low-temperature regime we expect the rank-one contributions (\ref{eq:nPointRank1}) to be dominant. We leave a more detailed study of this question for future work.

The most interesting question, however, is whether other simple theories of gravity in three dimensions -- such as general relativity -- should be regarded as an ensemble average over a space of CFTs.
About this we do not have much to say, except that in three dimensions perhaps the simplest conjecture is that: 
\be\label{conjecture}
Z_{\rm GR}(\Sigma) \equiv \int_{\partial M=\Sigma} Dg\, e^{-S[g]} = \sum_{\text{CFTs}~{\cal C}} \frac{1}{|{\rm Aut}({\cal C})|} Z_{\cal C}(\Sigma)
\ee
The left hand side of this expression is the Euclidean signature path integral of three dimensional general relativity with a negative cosmological constant, regarded as a function of the conformal structure and topology of the boundary $\Sigma$.  The right hand side is a sum over all two dimensional CFTs ${\cal C}$, with the most natural measure -- one over the cardinality of the symmetry group of the CFT -- inspired by the Siegel-Weil formula. We emphasize that both sides of this formula are extremely schematic, so an important challenge for the future is to understand the extent to which this formula can be made precise and either proven or disproven.\footnote{For example, it is not even clear that the central charge of the boundary CFTs should be related directly with the bulk cosmological constant via the usual Brown-Henneaux formula $c=\frac{3\ell}{2G}$. One could easily imagine that one should consider an ensemble where the central charge is allowed to vary, and  another variable (such as a chemical potential conjugate to $c$) is held fixed and related to the bulk cosmological constant. Similarly,  the space of CFTs includes both connected and discrete components, so the sum over CFTs in equation (\ref{conjecture}) should really include both a discrete sum as well as an integral over moduli.  We expect that the discrete sum should be more important since a typical CFT has no moduli, but of course we have not made this statement precise.}

\section*{Acknowledgements}

We are very grateful to N. Benjamin, H. Cohn, J. Cotler, S. Datta, S. Duary, A. Dymarsky, T. Hartman, K. Jensen, P. Kraus, P. Maity, A. Shapere, D. Stanford and especially E. Witten for helpful comments.
Research of AM is supported in part by the Simons Foundation Grant No. 385602 and the
Natural Sciences and Engineering Research Council of Canada (NSERC), funding reference number
SAPIN/00032-2015.  

\appendix

\section{Averaged density of states from modular crossing kernels}\label{app:crossingKernels}

\subsection{One-point function}\label{app:genusOneCrossingKernels}
In this appendix we will show that the averaged density of states in the Narain ensemble can be computed by summing over $SL(2,\mathbb{Z})$ modular crossing kernels.\footnote{We are grateful to Nathan Benjamin for discussions and collaboration regarding the material in this subsection.}

The holomorphic Virasoro characters for the $U(1)^D$ current algebra are given by
\begin{equation}
	\chi_h(\tau) = {q^h\over \eta(\tau)^D}.
\end{equation}
The Narain-averaged torus partition function is then defined through the following sum over $SL(2,\mathbb{Z})$ images of the vacuum character
\begin{equation}
\begin{aligned}
	\langle Z^p(\tau)\rangle &= |\eta(\tau)|^{4s}\sum_{\gamma\in\Gamma_\infty\backslash PSL(2,\ZZ)}|\chi_0(\gamma\tau)|^2\\
	&=  (\im\tau)^{-s}\sum_{\gamma\in\Gamma_\infty\backslash PSL(2,\ZZ)}{\im(\gamma\tau)^{s}}
\end{aligned}
\end{equation}
where $E_s$ is the real-analytic Eisenstein series, we have used the fact that $\sqrt{\im \tau}|\eta(\tau)|^2$ is modular invariant and recall that $s=\frac{D}{2}$. Here we will show that one can compute the density of states in this theory directly by making use of the generalized crossing kernels introduced in \cite{Benjamin:2020mfz}. These crossing kernels relate $SL(2,\ZZ)$ transforms of characters to integrals of characters in the original basis, and thus associate a contribution to the density of states to each term in the sum over geometries. The defining identity for the crossing kernel is
\begin{equation}
	(-i(c\tau+d))^{-w}e^{2\pi i(\gamma\tau)h} = \int_0^\infty dh'\, \mathbb{K}^{(w,\gamma)}_{h'h}e^{2\pi i \tau h'},
\end{equation}
where
\begin{equation}
	\mathbb{K}^{(w,\gamma)}_{h'h} = \varepsilon(w,\gamma)\left({2\pi\over c}\right)^{w}h'^{w-1}e^{{2\pi i\over c}(ah+dh')}{{}_0F_1\left(w;-{4\pi^2 h h'\over c^2}\right)\over \Gamma(w)},
\end{equation}
with the case of interest being $w=s$, where $\varepsilon(w,\gamma)$ is a weight-independent phase that will drop out in all applications of interest. The density of primary states is then given by
\begin{equation}
\begin{aligned}
	\langle\rho^p(h,\bar h)\rangle &= \sum_{\gamma\in \ZZ\backslash PSL(2,\mathbb{Z})}\mathbb{K}^{(s,\gamma)}_{h0}\overline{\mathbb{K}}^{(s,\gamma)}_{\bar h0}\\
	&= \sum_{n=-\infty}^\infty\sum_{c=1}^\infty\sum_{d'\in(\mathbb{Z}/c\mathbb{Z})^*}\left({2\pi\over c}\right)^{2s}{(h\bar h)^{s-1}e^{{2\pi i\over c}(d'+nc)(h-\bar h)}\over \Gamma(s)^2}\\
	&= \sum_{j=-\infty}^\infty \sum_{c=1}^\infty\sum_{d'\in(\mathbb{Z}/c\mathbb{Z})^*}\left({2\pi\over c}\right)^{2s}{(h\bar h)^{s-1}e^{{2\pi i\over c}d' j}\over \Gamma(s)^2}\delta(j-h+\bar h).
\end{aligned}
\end{equation}
The sum over $n$ localizes the density of states onto sectors of integer spin, so that we have\footnote{There is a factor of 2 here due to the Jacobian in writing the density of states in terms of $\Delta$ rather than $h$.}
\begin{equation}
\begin{aligned}
	\langle\rho^p_j(\Delta)\rangle &= \delta(\Delta)\delta_{j,0}+ \kappa_s(j){2\pi^{2s}(\Delta^2-j^2)^{s-1}\over \Gamma(s)^2}\\
	&= \begin{cases}
		\delta(\Delta)+{2\pi^{2s}\zeta(2s-1)\Delta^{2s-2}\over \Gamma(s)^2\zeta(2s)}, & j=0\\
		{2\pi^{2s}\sigma_{2s-1}(j)(\Delta^2-j^2)^{s-1}\over |j|^{2s-1}\Gamma(s)^2\zeta(2s)}, & j\ne 0
	\end{cases}
\end{aligned}
\end{equation}
precisely reproducing the result of the Siegel-Weil formula as written for instance in \cite{Afkhami-Jeddi:2020ezh}. The canonical ensemble primary partition function in a sector of fixed spin $j$ is then given by integrating against the characters, with the following result familiar from the standard Fourier decomposition of the Eisenstein series
\begin{equation}
\begin{aligned}
	\langle Z^p_j(y)\rangle &= \begin{cases}
		1 + {\sqrt{\pi}\zeta(2s-1)\Gamma(s-1/2)\over \Gamma(s)\zeta(2s)}y^{1-2s}, & j=0\\
		{2\pi^s\sigma_{2s-1}(j)\over (|j| y)^{s-\half}\Gamma(s)\zeta(2s)}K_{s-\half}(2\pi |j| y), & j\ne 0,
	\end{cases}
\end{aligned}
\end{equation}
where
\begin{equation}
\langle Z^p(\tau)\rangle = \sum_{j\in\mathbb{Z}}e^{2\pi i j x}\langle Z^p_j(y)\rangle.
\end{equation} 

\subsection{Two-point function from a genus-two crossing kernel}\label{app:genusTwoCrossingKernels}
In this section we will show that the rank-two part of the two-point function of the density of states can also be straightforwardly computed by summing over a certain genus-two crossing kernel. The rank-two part of the two-point function can be written in terms of a sum over $2\times 2$ rational symmetric matrices
\begin{equation}
	\langle Z^p(\tau_1)Z^p(\tau_2)\rangle_2 = \sum_{\substack{P = P^T \\ \rm rational}}\nu(P)^{-2s}|\det(\Omega+P)|^{-2s},
\end{equation}
where $\Omega = \diag(\tau_1,\tau_2)$ packages the modular parameters of the boundary tori and in what follows we will write $P = \mat{p_1}{p}{p}{p_2}$. To proceed, we write the determinant appearing in the summand in terms of the following integral
\begin{equation}
\begin{aligned}
	\det(\Omega+P)^{-s} &= \left((p_1+\tau_1)(p_2+\tau_2)-p^2\right)^{-s}\\
	&= \int_0^\infty dh_1\int_0^\infty dh_2 \, \mathbb{K}_{h_1h_2;\rm vac}^{(s,P)} e^{2\pi i \tau_1 h_1}e^{2\pi i \tau_2 h_2},
\end{aligned}
\end{equation} 
where $\mathbb{K}$ is a sort of genus-two crossing kernel (in the sense that it relates terms that appear in the $Sp(4,\mathbb{Z})$ modular sum to a density of states in the original channel) given by
\begin{equation}
	\mathbb{K}_{h_1h_2;\rm vac}^{(s,P)} = \varepsilon(s,P) {4\pi^{2s}\over \Gamma(s)^2}\left(4h_1h_2\right)^{s-1}e^{2\pi i (p_1h_1+p_2h_2)}\ofone\left(s;-4\pi^2 p^2 h_1h_2\right).
\end{equation}
As before, $\varepsilon$ is a phase independent of the weights that will cancel when we combine the holomorphic and anti-holomorphic sectors. 

We can then write the full rank-two density-density correlator in terms of a sum over this $Sp(4,\mathbb{Z})$ crossing kernel
\begin{equation}
\begin{aligned}
	\langle\rho^p(h_1,\bar h_1)\rho^p(h_2,\bar h_2)\rangle_2 =& \, \sum_{\substack{P = P^T \\ \rm rational}}\nu(P)^{-2s} \mathbb{K}_{h_1h_2;\rm vac}^{(s,P)} \overline{\mathbb{K}}_{\bar h_1\bar h_2;\rm vac}^{(s,P)}\\
	=& \sum_{\substack{R = R^T \\ \text{rational mod 1}}}\sum_{\substack{N = N^T \\ \rm integral}}\nu(R)^{-2s} {16\pi^{4s}\over \Gamma(s)^4}(16 h_1 \bar h_1 h_2 \bar h_2)^{s-1} e^{2\pi i\left((h_1-\bar h_1)(r_1+n_1)+(h_2-\bar h_2)(r_2+n_2)\right)}\\
	& \, \times \ofone\left(s;-4\pi^2 (r+n)^2 h_1 h_2\right)\ofone\left(s;-4\pi^2(r+n)^2 \bar h_1 \bar h_2\right)\\
	=& \, \sum_{j_1,j_2,n\in\mathbb{Z}}\sum_{r_1,r_2,r\in\mathbb{Q}/\mathbb{Z}}\delta(j_1-h_1+\bar h_1)\delta(j_2-h_2+\bar h_2)\, \nu(R)^{-2s}e^{2\pi i (j_1 r_1+j_2 r_2)}\\
	& \, \times {16\pi^{4s}\over\Gamma(s)^4}(16 h_1\bar h_1 h_2 \bar h_2)^{s-1}\ofone\left(s;-4\pi^2(r+n)^2h_1h_2\right)\ofone\left(s;-4\pi^2(r+n)^2\bar h_1 \bar h_2\right).
	\end{aligned}
\end{equation}
In the last line we realized the sums over $n_1, n_2$ as Dirac combs that project onto sectors of definite integer spins $j_1, j_2$. From this it is easy to read off the full two-point function of the density of states in particular spin sectors. Up to a factor of 4 from to the Jacobian in going between the weights $h_1,h_2$ and the dimensions $\Delta_1,\Delta_2$, this is precisely the result (\ref{eq:rankTwoDensityDensity2}).

\section{Higher-degree Eisenstein series}\label{app:eisensteinFourier}
In this section we will review some salient features of the modular group of degree $n$, relevant for the construction and Fourier decomposition of the higher degree Eisenstein series that arise as observables in the averaged Narain lattice CFTs. We will very closely follow the discussion in chapters $11, 12$ and $18$ of \cite{Maa__1971}.

\subsection{The degree $n$ modular group}
The degree-$n$ Eisenstein series is defined on the Siegel upper half-space $\mathcal{H}_n$ of complex, symmetric $n\times n$ matrices with positive-definite imaginary part
\begin{equation}
	\mathcal{H}_n = \{\Omega_{ij}\in\mathbb{C} \,|\, \Omega = \Omega^T,~\im\Omega>0\}.
\end{equation}
We will mostly be interested in the case that $\Omega$ is the period matrix of a Riemann surface of genus $n$, but the Eisenstein series is well-defined on the larger space $\mathcal{H}_n$. The degree $n$ modular group acts on $\Omega$ by
\begin{equation}\label{eq:symplecticAction}
	\gamma\Omega = (A\Omega+B)(C\Omega+D)^{-1},~\gamma = \mat{A}{B}{C}{D}\in Sp(2n,\ZZ).
\end{equation}

 The Eisenstein series is defined by summing over images of $\det\im\Omega^s$ in the modular group
\begin{equation}\label{eq:higherGenusEisenstein}
	E_s^{(n)}(\Omega) = \sum_{\gamma\in P\backslash Sp(2n,\ZZ)}(\det\im\gamma\Omega)^s = (\det\im\Omega)^s \sum_{\{C,D\}}|\det(C\Omega+D)|^{-2s},
\end{equation}
where the sum over the pair of matrices $(C,D)$ is over a complete set of equivalence classes of symmetric coprime matrices as we will shortly describe. The quotient by $P$ is by the parabolic subgroup
\begin{equation}\label{eq:parabolicSubgroup}
P = \left\{\mat{A}{B}{0}{D}\in Sp(2n,\ZZ)\right\}
\end{equation} 
that fixes $|\det\im\Omega|$.  

We will proceed by describing in more detail the modular group of degree $n$. We will start by introducing the square matrix $I$ of order $2n$
\begin{equation}
	I = \mat{0}{\id}{-\id}{0},
\end{equation}
where $\id$ is the identity matrix of order $n$. The symplectic group $Sp(2n,\RR)$ is the set of real-valued matrices $\gamma$ such that $I[\gamma] \equiv \gamma^T I \gamma = I$. If we write $\gamma$ in terms of real $n\times n$ matrices $A,B,C,D$ as in (\ref{eq:symplecticAction}) then this condition amounts to
\begin{equation}
\begin{aligned}\label{eq:symplecticConditions}
	A D^T - B C^T &= \id\\
	AB^T - BA^T &= 0\\
	CD^T - DC^T &= 0.
\end{aligned}	
\end{equation}
The modular group of degree $n$, $Sp(2n,\ZZ)$, is the set of matrices $\gamma$ as in (\ref{eq:symplecticAction}) subject to (\ref{eq:symplecticConditions}) with the further restriction that the constituent matrices $A,B,C,D$ are integer-valued. 

In what follows we will characterize the space of the pair of square $n\times n$ integer-valued matrices $(C,D)$ that could serve as the lower row of an element of the modular group. It turns out that matrices $(C,D)$ can comprise the lower row of an element of the modular group $Sp(2n,\ZZ)$ if and only if they form what is known as a symmetric coprime pair (generalizing the fact that in the case $n=1$, the elements of the lower row of a matrix in $SL(2,\ZZ)$ must be coprime integers). From (\ref{eq:symplecticConditions}) we see that $(C,D)$ must be a \emph{symmetric pair}, namely $CD^T=DC^T$. The pair $(C,D)$ is deemed \emph{coprime} in the sense that if $GC,\,GD$ are both square integral matrices, then $G$ must be as well. It is relatively straightforward to see that a matrix $(C,D)$ coming from the lower row of a modular group element must be symmetric coprime matrices, and in \cite{Maa__1971} it is shown how to construct a modular group element from a given coprime pair. 

Moreover, due to the quotient by the parabolic subgroup $P$ defined in (\ref{eq:parabolicSubgroup}) in the modular sum involved in the Eisenstein series, in what follows two symmetric pairs of coprime matrices $(C_1,D_1)$ and $(C_2,D_2)$ are to be regarded as equivalent or ``associated'' if they are related by the action of a unimodular matrix\footnote{A unimodular matrix $M$ is a square integral matrix with determinant $\det M = \pm 1$. Such matrices have an inverse that is also an integral matrix.} $U$
\begin{equation}
	(C,D) \sim U(C,D),\quad\text{$U$ unimodular}.
\end{equation}
Indeed, it is easy to see that such coprime pairs give identical contributions to the Eisenstein series (\ref{eq:higherGenusEisenstein}). One can show that two such pairs are associated if and only if
\begin{equation}
	C_1 D_2^T = D_1 C_2^T.
\end{equation}
We will denote by $\{C,D\}$ this class of coprime symmetric pairs of matrices associated to $(C,D)$. 

In the modular sum involved in the higher-degree Eisenstein series, one sums over one representative per class. Such equivalence classes are organized by the rank of the matrix $C$. The case that $\rank C = 0$ is trivial, as we can take $(C,D) = (0,\id)$ as a representative, so in what follows we will assume $\rank C = r$ with $0<r\le n$. We would like to identify a particular simple representative for each class in order to make the sum involved in the Eisenstein series more concrete. Here we will briefly review the construction of \cite{Maa__1971}, where it is shown that for each rank $r$ we can rephrase the sum over representatives as a sum over $r\times r$ matrices $C^{(r)}$ and $D^{(r)}$ with $\det C^{(r)} \ne 0$, augmented by a sum over primitive $n\times r$ matrices $Q^{(n,r)}$. 

We start by seeking two unimodular $n\times n$ matrices $U_1,U_2$ such that \cite{Maa__1971}
\begin{equation}\label{eq:ceeMatrix}
	C = U_1^{-1}\mat{C_1}{0}{0}{0}U_2^T,
\end{equation}
where $C_1$ is an $r\times r$ matrix with $\det C_1 \ne 0$, and
\begin{equation}\label{eq:deeMatrix}
	D = U_1^{-1}\mat{D_1}{D_2}{D_3}{D_4}U_2^{-1},
\end{equation}
where $D_1$ is an $r\times r$ matrix. If $(C,D)$ is a symmetric pair, then so is $U_1(C,D)$, and from the symmetry property (\ref{eq:symplecticConditions}) of $U_1(C,D)$ one can show that $(C_1,D_1)$ also forms a symmetric pair and deduce that $D_3 = 0$. By making the replacement
\begin{equation}
	U_1 \to \mat{\id}{D_2}{0}{D_4}U_1,
\end{equation}
(\ref{eq:ceeMatrix}) and (\ref{eq:deeMatrix}) simplify to
\begin{equation}
\begin{aligned}
	U_1 C &\to \mat{C_1}{0}{0}{0}U_2^T\\
	U_1 D &\to \mat{D_1}{0}{0}{\id}U_2^{-1} 
\end{aligned}
\end{equation}
It follows from this construction that the pair $(C_1,D_1)$ is also necessarily coprime \cite{Maa__1971}.

Furthermore, if one acts on $U_2$ from the right by a unimodular matrix that only has a nontrivial upper-left $r\times r$ block, for example
\begin{equation}
	U_2\to U_2\mat{U_3}{0}{0}{\id},
\end{equation}
where $U_3$ is an $r\times r$ unimodular matrix, then one can simply make the replacement $(C_1,D_1)\to (C_1U_3^T,D_1 U_3^{-1})$ and nothing is changed. To take account of this equivalence, let us introduce the $n\times r$ matrix $Q^{(n,r)}$, formed by taking the first $r$ columns of $U_2$. We are thus motivated to introduce the equivalence class $\{Q^{(n,r)}\}$
\begin{equation}
	Q^{(n,r)}\sim Q^{(n,r)}U_3,~\text{for $U_3$ $r\times r$ unimodular}
\end{equation}
of primitive $n\times r$ matrices that can be completed to a unimodular $n\times n$ matrix (that we refer to as $U_2$ above). Importantly, in the maximal-rank case of $r=n$ we have only $Q^{(n,n)} = \id$.

Finally, by replacing
\begin{equation}
	U_1 \to \mat{U_4}{0}{0}{\id}U_1
\end{equation}
with $U_4$ an $r\times r$ unimodular matrix, we have $(C_1,D_1)\to U_4(C_1,D_1)$ and so these pairs are also associated. Thus one is free to choose $(C_1,D_1)$ as any representative of the class $\{C_1,D_1\}$.

The upshot of this construction is that one can associate to each equivalence class $\{C,D\}$ where $\rank C = r$ the equivalence classes of $r\times r$ matrices $\{C^{(r)},D^{(r)}\}$ with $\det C^{(r)} \ne 0$, augmented by the equivalence class of primitive matrices $\{Q^{(n,r)}\}$, such that
\begin{equation}
\begin{aligned}
	C &= \mat{C^{(r)}}{0}{0}{0}U^T\\
	D &= \mat{D^{(r)}}{0}{0}{\id}U^{-1}\\
	U &= (Q^{(n,r)},*),
\end{aligned}
\end{equation}
where $U$ is the completion of $Q^{(n,r)}$ into a unimodular matrix.  In \cite{Maa__1971} it is shown that this association is in fact one-to-one. In fact, since $C^{(r)}$ is invertible, one can further define the matrix $R = (C^{(r)})^{-1}D^{(r)}$, and in \cite{Maa__1971} it is shown that the set of all equivalence classes $\{C^{(r)},D^{(r)}\}$ with $\det C^{(r)} \ne 0$ is precisely the set of all rational symmetric $r\times r$ matrices $R$, and the map is one-to-one.

\subsection{Fourier decomposition of the higher-degree Eisenstein series}
To begin unpackaging the sum in the definition of the Eisenstein series, recall from the previous subsection the one-to-one correspondence between the equivalence class $\{C,D\}$ and that of $r\times r$ matrices $\{C^{(r)},D^{(r)}\}$ where $r=\rank C$, $\det C^{(r)}\ne 0$ and $\{Q^{(n,r)}\}$ of primitive $n\times r$ matrices that can be completed to a unimodular $n\times n$ matrix. Then, since in the case that $\rank C = r$ 
\begin{equation}
	\det(C\Omega+D) = \pm \det (C^{(r)} \Omega[Q^{(n,r)}]+D^{(r)})
\end{equation}
we can rewrite (\ref{eq:higherGenusEisenstein}) as
\begin{equation}\label{eq:higherGenusEisenstein2}
	{E^{(n)}_s(\Omega)\over (\det\im \Omega)^s} = 1+\sum_{r=1}^n\sum_{\{C^{(r)},D^{(r)}\}}\sum_{\{Q^{(n,r)}\}}\det(C^{(r)})^{-2s}|\det(\Omega[Q^{(n,r)}]+(C^{(r)})^{-1}D^{(r)})|^{-2s}.
\end{equation}

Recall that the sum over $\{C^{(r)},D^{(r)}\}$ can be replaced by a sum over symmetric rational $r\times r$ matrices $R = (C^{(r)})^{-1}D^{(r)}$. Then one can find $r\times r$ unimodular matrices $U,V$ such that
\begin{equation}
	U^{-1}RV = \diag({p_i/q_i}),~\text{where $(p_i,q_i)=1$ and ${p_i/ q_i}\ge 0$.}
\end{equation}
The ratios $p_i/q_i$ are called the elementary divisors of $R$. Then one can write the determinant of $C^{(r)}$ in terms of the product of the denominators of the elementary divisors of $R$ \cite{Maa__1971}
\begin{equation}
	\nu(R) \equiv \det C^{(r)} = \prod_{i=1}^r q_i.
\end{equation}
Finally, this allows us to rewrite the Eisenstein series in terms of a sum over symmetric rational matrices
\begin{equation}
\begin{aligned}\label{eq:higherGenusEIsenstein3}
	{E^{(n)}_s(\Omega)\over (\det\im \Omega)^s} &= 1+\sum_{r=1}^n\sum_{\{Q^{(n,r)}\}}\sum_{\substack{P=P^T \\ \rm rational}}\nu(P)^{-2s}|\det(\Omega[Q^{(n,r)}]+P)|^{-2s}\\
	&= 1 + \sum_{r=1}^n\sum_{\{Q^{(n,r)}\}}\sum_{\substack{R=R^T \\ {\rm rational} \mymod 1}}\sum_{\substack{N = N^T \\ {\rm integral}}}\nu(R)^{-2s}|\det(\Omega[Q^{(n,r)}]+R+N)|^{-2s},
\end{aligned}
\end{equation}
where in the second line we wrote $P = R+N$ where $R$ is a symmetric matrix with entries that are rational mod one, and $N$ is a symmetric integral matrix. 

To proceed we define the function
\begin{equation}
	\psi_{s,r}(\Omega) = \sum_{\substack{N=N^T\\ {\rm integral}}}|\det(\Omega+N)|^{-2s},
\end{equation}
where $\Omega$ and $N$ are understood to be $r\times r$ matrices. 
By construction this function is invariant under integral shifts of the real part of $\Omega \equiv X + iY$, $X\to X+N$ for $N$ symmetric and integral, and thus admits a Fourier decomposition
\begin{equation}
	\psi_{s,r}(\Omega) = \sum_{\substack{J = J^T \\ {\rm semi-integral}}}\widehat\psi_{s,r}(Y,J)e^{2\pi i\sigma(JX)},
\end{equation}
where, for symmetric $r\times r$ matrices $A,B$, $\sigma$ is defined by the trace
\begin{equation}
	\sigma(AB) = \tr(AB) =  \sum_{i=1}^r A_{ii}B_{ii} + \sum_{i< j}^r2A_{ij}B_{ij}.
\end{equation}
The condition that $J$ be symmetric and semi-integral requires
\begin{equation}
\begin{aligned}
	J_{ii} &\in \mathbb{Z}\\
	J_{ij} &= J_{ji} \in \half\mathbb{Z},~i < j.
\end{aligned}
\end{equation}
The Fourier coefficients $\widehat\psi_{s,r}$ can be written in terms of an integral over the unit $X$ cube $\mathcal{X}$
\begin{equation}
\begin{aligned}
	\widehat\psi_{s,r}(Y,J) &= \int_{\mathcal{X}}[dX]\,\psi_{s,r}(\Omega)e^{-2\pi i \sigma(JX)}\\
	&= \int_{-\infty}^\infty[dX]\, |\det\Omega|^{-2s}e^{-2\pi i \sigma(JX)},
\end{aligned}
\end{equation}
where $[dX] = \prod_{i\le j}dX_{ij}$ is the flat measure and in the second line we used the explicit form of $\psi_s$ as a sum over integral shifts of $\Omega$ and the invariance of the exponential factor under such shifts. After some standard manipulations, one can show that \cite{Maa__1971}
\begin{equation}
	\widehat\psi_{s,r}(Y,J) = {2^r \pi^{2rs}\over \Gamma_r(s)^2}h_{s,r}(Y,J),
\end{equation}
where
\begin{equation}\label{eq:higherGenusFourierIntegral}
\begin{aligned}
	h_{s,r}(Y,J) &= \int_{\Delta \pm J > 0}[d\Delta]\, e^{-2\pi \sigma(\Delta Y)}\det(\Delta+J)^{s-\half(r+1)}\det(\Delta-J)^{s-\half(r+1)}\\
	\Gamma_r(s) &= \pi^{{1\over 4}r(r-1)}\prod_{j=0}^{r-1}\Gamma(s-j/2).
\end{aligned}
\end{equation}
The key to this rewriting is the following identity
\begin{equation}
	\int_{H>0}[dH]\,e^{i\sigma(H\Omega)}\det(H)^{s-{r+1\over 2}} = \Gamma_r(s)\det(-i\Omega)^{-s},
\end{equation}
where the integral is over symmetric positive-definite $r\times r$ matrices $H$.
The functions $h_{s,r}(Y,J)$ are sometimes known as confluent hypergeometric functions on tube domains \cite{Shimura_1982}, and it is by studying properties of these functions that the functional equation (\ref{eq:eisensteinFunctionalEquation}) satisfied by the meromorphic continuation of the higher-degree Eisenstein series was established. 

Thus the Fourier decomposition of the degree $n$ Eisenstein series (\ref{eq:higherGenusEIsenstein3}) can be written as
\begin{equation}\label{eq:higherGenusFourier}
\begin{aligned}
	{E^{(n)}_s(\Omega)\over (\det\im \Omega)^s} &= 1 + \sum_{r=1}^n \sum_{\substack{J^{(r)} = (J^{(r)})^T\\ {\rm semi-integral}}}\sum_{\{Q^{(n,r)}\}}S_{s,r}(J)\widehat{\psi}_{s,r}(Y[Q],J)e^{2\pi i \sigma(X[Q]J)}\\
	 &= 1 + \sum_{r=1}^n \sum_{\substack{J^{(r)} = (J^{(r)})^T\\ {\rm semi-integral}}}\sum_{\{Q^{(n,r)}\}}{2^r\pi^{2rs}\over \Gamma_r(s)^2}S_{s,r}(J)h_{s,r}(Y[Q],J)e^{2\pi i \sigma(X[Q]J)},
\end{aligned}
\end{equation}
where $S_{s,r}(J)$ is Siegel's ``singular series''
\begin{equation}\label{eq:SiegelSingularSeries}
	S_{s,r}(J) = \sum_{\substack{R=R^T \\ {\rm rational} \mymod 1}}\nu(R)^{-2s}e^{2\pi i \sigma(JR)},
\end{equation}
and in the summand on the right-hand side we have omitted for the sake of brevity the superscripts on the matrices $Q^{(n,r)}$ and $J^{(r)}$ denoting their sizes. Some special values of the singular series are known explicitly, for example in the case that $J$ is the zero matrix we have \cite{Maa__1971}
\begin{equation}
	S_{s,r}(0) = {\zeta(2s-r)\over \zeta(2s)}\prod_{j=1}^r {\zeta(4s-r-j)\over \zeta(4s-2j)}.
\end{equation}

The sum (\ref{eq:higherGenusEisenstein}) converges provided $\re s>{n+1\over 2}$. Like the genus-one Eisenstein series, it can be formally defined by analytic continuation in the complex $s$ plane away from this domain, but the argument relating it to a partition function of an averaged Narain lattice CFT established in \cite{Maloney:2020nni} only holds for half-integer $s>{n+1\over 2}$. The meromorphic continuation of the Eisenstein series can be shown to satisfy the functional equation \cite{Kaufhold_1959,Shimura_1982,Mizumoto_1993,Kalinin_1978}
\begin{equation}\label{eq:eisensteinFunctionalEquation}
  \mathcal{E}_s^{(n)}(\Omega) = \mathcal{E}^{(n)}_{{n+1\over 2}-s}(\Omega)
\end{equation}
where 
\begin{equation}
  \mathcal{E}^{(n)}_s(\Omega) = \Lambda(s)\left(\prod_{j=1}^{\floor{{n\over 2}}}\Lambda(2s-j)\right)E^{(n)}_s(\Omega),\quad \Lambda(s) = \pi^{-s}\Gamma(s)\zeta(2s).
\end{equation}
The function $\mathcal{E}^{(n)}_s(\Omega)$ can be shown to have poles of finite order at $s = {k\over 4}$ for $k = 0, 1, \ldots, 2n+2$, including a simple pole at $ s= {n+1\over 2}$ with residue $\prod_{j=1}^{\floor{{n\over 2}}}\Lambda\left(j+\half\right)$ \cite{Kalinin_1978}.

\bibliographystyle{JHEP}
\bibliography{narainSFF.bib}
\end{document}